\documentclass[a4paper,11pt]{article}
\usepackage{jheppub}

\usepackage{epsfig}
\usepackage{amssymb,amsmath}
\usepackage{pstricks}
\usepackage{pst-all}
\usepackage[utf8]{inputenc}

\newcommand{\eq}{\begin{equation}}
\newcommand{\qe}{\end{equation}}
\newcommand{\eqa}{\begin{eqnarray}}
\newcommand{\qea}{\end{eqnarray}}
\newcommand{\tauint}{{\tau_\mathrm{int}}}
\newcommand{\tauintreal}{{\tau_\mathrm{int}^\mathrm{real}}}

\newcommand{\de}{\mathrm{d}\hspace{0.01em}}

\title{Thermodynamics of the O(3) model in 1+1 dimensions: lattice vs.~analytical results}

\author[a]{Elina Seel,}
\author[abc]{Dominik Smith,}
\author[a]{Stefano Lottini,}
\author[a]{Francesco Giacosa}
\affiliation[a]{Institut f\"ur Theoretische Physik, Johann Wolfgang Goethe-Universit\"at, \\
	Max-von-Laue-Str.~1, 60438 Frankfurt am Main, Germany}
\affiliation[b]{Technische Universit\"{a}t Darmstadt, Institut f\"{u}r Kernphysik, \\
	Schlossgartenstr.~2, 64289 Darmstadt, Germany}
\affiliation[c]{Fakult\"{a}t f\"{u}r Physik, Universit\"{a}t Bielefeld, 33615 Bielefeld, Germany}
\emailAdd{seel, smith, lottini, giacosa@th.physik.uni-frankfurt.de}

\abstract{A detailed study of the thermodynamics of the $O(N=3)$ model in 1+1
dimensions is presented, employing a two-particle-irreducible resummation prescription as well as
fully nonperturbative finite-temperature lattice simulations. The analytical
results are computed using the Cornwall-Jackiw-Tomboulis (CJT) formalism and the auxiliary field method to
one- and to two-loop order. The lattice results are obtained through Monte Carlo simulation
for various lattice spacings. The
analytical and lattice results for pressure, trace anomaly, and energy density,
resembling closely those of four-dimensional Yang-Mills theories,
are compared with each other.
We find that to one-loop
order there is a good correspondence between the CJT formalism and the lattice
study for low temperatures.
However, at high $T$ the two-loop calculation fares better, correcting
for the overestimation from the former approximation.
}

\begin{document}
\maketitle

\section{Introduction}
\label{sec:intro}

As discussed by Polyakov \cite{polyakov_1}, the two-dimensional $O(N)$
nonlinear sigma model has interesting features in common with four-dimensional
non-Abelian gauge theories. For these reasons it has been investigated in the
past as a suitable toy model for QCD, see Refs.~\cite{novikov,rim, flyv,
gross, warringa1, warringa2} and refs.~therein.
Remarkably, the two-dimensional $O(3)$ model emerges also naturally as a
particular subsystem of four-dimensional theories: see Ref.~\cite{Shifman:2010id}, in
which the dual Meissner mechanism for confinement is investigated in the
framework of supersymmetric QCD.

In $1+1$ dimensions, the coupling constant is dimensionless, the theory is
renormalisable both perturbatively and in the $1/N$ expansion, and has also a
negative beta-function, thus showing asymptotic freedom \cite{polyakov_1}.
Moreover, the two-dimensional $O(N)$ model is conformally invariant at the
classical level, but, just as Yang-Mills theories in four dimensions, at the
quantum level an energy scale emerges due to quantum corrections (trace
anomaly). In fact, it exhibits a nonperturbative mass gap $m=\mu\exp(-2\pi/g^{2})$
generated dynamically due to the interactions, where $g$ is the
coupling constant and $\mu$ the renormalisation parameter. Note that: (i)
since the mass is non-analytic for $g\rightarrow0$, it vanishes in
perturbation theory; (ii) unlike the case of more than two dimensions, here no
spontaneous breaking of the global $O(N)$ symmetry takes place: this is
related to the Mermin-Wagner-Coleman theorem \cite{mermin, coleman},
which forbids spontaneous breaking of a continuous
global symmetry in a $(1+1)$-dimensional homogeneous system at any temperature.

The effective potential at zero temperature has been investigated in Refs.
\cite{root,biscari}. In Ref.~\cite{dine_fischler} the model was studied at
finite temperature in perturbation theory and to leading order (LO) in the
large-$N$ limit; this was subsequently extended in
Refs.~\cite{warringa1, warringa2} to next-to-leading order (NLO) in
the $1/N$ expansion. It is found that the NLO effective potential contains
temperature-dependent divergences. Fortunately, it turns out that the
effective potential can still be renormalised at the minimum.

In this work we study the nonlinear $O(N=3)$ model in 1+1 dimensions at
nonzero temperature both analytically and numerically.
On the analytical
side, we compute the thermodynamic properties by employing the
Cornwall-Jackiw-Tomboulis (CJT) formalism \cite{CJT} and using the auxiliary field
method in the one- and two-loop approximation. We present analytical
expressions for the renormalised pressure, the trace anomaly, the energy
density, and the quasi-particle mass of the scalar fields.
On the numerical side, we compute
the thermodynamic quantities via ``first-principles''
Monte Carlo lattice calculation using the
so called ``integral method''. We then compare the analytic and the lattice results: at one
loop, a good agreement is found for small temperatures, but the analytic result
for the pressure is too large when $T$ increases. When going to the two-loop
approximation, a considerable improvement in the high-$T$ domain is obtained.

Among the motivations of the present study are the following. (i) A check of the
validity of the CJT formalism: through the direct comparison with our
precise lattice results we test to which extent the widely used CJT
theoretical approach gives reliable results for thermodynamic quantities
such as pressure and energy density. No free parameter can be adjusted when
doing this comparison. (ii) Lattice evaluation: the two-dimensional $O(N=3)$
model has been already studied in a variety of lattice-related contexts 
(see Section \ref{sec:montecarlo} for references), but to our
knowledge a systematic evaluation of all thermodynamic quantities was not
yet presented. The lattice results are interesting on their own, and can be
used by other groups who study this system. (iii) Due to the mentioned
similarity between the two-dimensional $O(N)$ model and gauge theories, we aim
at a better understanding of fundamental issues such as the nonperturbative mass
generation through the trace anomaly and its behaviour when the temperature
is raised: the link between a non-perturbative vacuum and a perturbative
high-temperature domain is non-trivial.

A further interesting motivation for the study of the present model is that,
for the choice $N=3$, the model exhibits instantonic solutions both at zero
\cite{novikov} and nonzero temperature \cite{bruk1, bruk2}. In fact, the
$O(3)$ model in $1+1$ dimensions is topologically equivalent the $CP^{1}$
model, whereas $CP^{n-1}$ theories admit instantons for each value of $n$
\cite{eichenherr,banerjee}. The instanton solutions are~topologically
nontrivial field configurations localised in space and time, associated to
classical, finite-action solutions of Euclidean field equations. They give
rise to tunneling processes between different classical vacua
\cite{instantons_1}. For this reason they have been regarded in the framework
of Yang-Mills theories $SU(N>1)$ (see for instance \cite{gross,calorons_1,calorons_2,calorons_3}
and refs.~therein) as a possible crucial ingredient for understanding
nonperturbative properties, such as confinement and the emergence of a mass gap. 

Finally, one may wonder what is the role of instantons and calorons in the thermodynamics
of this model; although our analytical method does not
distinguish topologically trivial and nontrivial field configurations,
we suspect, in agreement with \cite{warringalast},
that the role of instantons and/or calorons can be relevant.

The manuscript is organised as follows: in Sec.~\ref{sec:analytical} we present the model,
derive the effective potential to one- and two-loop order, and derive the
analytic expressions for pressure, trace anomaly and energy
density. In Sec.~\ref{sec:montecarlo} we present a detailed lattice study of
the model. In Sec.~\ref{sec:comparison} we show and discuss the
comparison of analytic and lattice results. The reader interested only in the
results may skip the details of
Secs.~\ref{sec:analytical} and \ref{sec:montecarlo} and go
directly to Sec.~\ref{sec:comparison}.
Finally, conclusions and outlooks are described in Sec.~\ref{sec:conclusions}.

Throughout the paper, we will use natural units $\hbar=c=k_{B} \equiv 1$. All
computations at nonzero temperature $T$ are done in the imaginary-time
formalism, with the shorthand
\eq
\int_{K}f(K) \equiv T\sum_{n=-\infty}^{+\infty}\int\frac{\de k}{\left(  2\pi\right)
}f(k,2\pi nT)\;\;,
\qe
where we have used the Euclidean momentum $K=(k_{x}=k,k_{\tau}).$

\section{The model and the analytic results}
\label{sec:analytical}

\subsection{Generating functional}

The nonlinear $O(N)$ model in $1+1$ dimensions at nonzero temperature is
defined by the following generating functional:
\eq
\ \mathcal{Z}=\mathcal{N}\int\mathcal{D}\Phi~\delta\left[  \Phi^{2}-\frac
{N}{g^{2}}\right]  \exp\left(  -\overset{\ }{\int_{0}^{1/T}}\de\tau
\int_{-\infty}^{\ \infty}\de x~\mathcal{L}_{0}\right)  \;\;,\label{z}
\qe
where $g$ is the dimensionless coupling constant. The $N$ fields of the model
are contained in the column matrix $\Phi$, which for future purposes we write
as
\eq
\Phi^{t}=\left(  \sigma,\pi_{1},\ldots,\pi_{N-1}\right)\;\;.
\qe
Moreover, we also use the notation $\Phi^{2}=\Phi^{t}\Phi$ and $\vec{\pi}
^{t}=\left(  \pi_{1},\ldots,\pi_{N-1}\right)  .$

The Lagrangian $\mathcal{L}_{0}$ entering Eq.~(\ref{z}) is a simple
Lagrangian for $N$ massless scalar fields, which in Euclidean space reads:
\eq
\mathcal{L}_{0}= \tfrac{1}{2}(\partial_{\mu}\Phi)^{2}
\;\;,
\qe
with $\partial_{1}=\partial_{x},$ $\partial_{2}=\partial_{\tau}$.

Due to the $\delta$-function entering the partition function (\ref{z}) the
fields of the model are constrained by the condition
\eq
\Phi^{2}=\frac{N}{g^{2}}\;\;.
\label{constraint}
\qe
This nonlinear constraint enforces the $N$ degrees of freedom of the model on an
($N-1$)-dimensional hypersphere and induces interactions between them. Using
the mathematically well-defined (i.e.~convergent) form of the usual
representation of the functional $\delta$-function
\eq
	\delta\left[ \Phi^{2}-\frac{N}{g^{2}}\right]  \sim
	\lim_{\varepsilon\rightarrow 0^{+}}\int\mathcal{D}\alpha~
	\exp {\left\{  -\ \overset{\ }{\int_{0}^{1/T}}\de\tau\int_{-\infty}^{+\infty
	}\de x\left[  \frac{i\alpha}{2}\left(  \Phi^{2}-\frac{N}{g^{2}}\right)
		+\frac{N\varepsilon}{8}\alpha^{2}
	\right]  \right\}  }\;\;,
\qe
the generating functional and the corresponding Lagrangian can be rewritten as
follows:
\eq
	\mathcal{Z}=\lim_{\varepsilon\rightarrow\;\;0^{+}}\mathcal{N}'
	\int\mathcal{D}\alpha~\mathcal{D}\Phi~\exp\left(  -\overset{\ }{\int_{0}^{1/T}}
	\de\tau\int_{-\infty}^{\ \infty}\de x\mathcal{L}\right)\;\;,
\label{zalfa}
\qe
where
\eqa
	\mathcal{L} &=& \tfrac{1}{2}(\partial_{\mu}\Phi)^{2}
		+U(\Phi,\alpha)\;\;;\nonumber\\
	U(\Phi,\alpha) &=& \dfrac{i}{2}\alpha\left(  \Phi^{2}-\dfrac{N}{g^{2}}\right)
		+\dfrac{N\varepsilon}{8}\alpha^{2}\;\;.
	\label{lalfa}
\qea
In the last expression $\alpha$ is an unphysical auxiliary field serving as a
Lagrange multiplier. In the next Subsections we shall use the form in
Eqs.~(\ref{zalfa}) and (\ref{lalfa}) for our explicit calculation of the
thermodynamical properties. Interestingly, by further eliminating $\alpha$ we
obtain the following equivalent form of the generating functional:
\eq
\mathcal{Z}=\lim_{\varepsilon\rightarrow 0^{+}}\widetilde{\mathcal{N}}
\int\mathcal{D}\Phi \exp  \left\{  -\ \overset{\ }{\int_{0}^{1/T}}\de\tau
\int_{-\infty}^{\ \infty}\de x\left[  \tfrac{1}{2}(\partial_{\mu}\Phi)^{2}
+\frac{\left(  \Phi^{2}-N/g^{2}\right)  ^{2}}{2 N \varepsilon}\right]  \right\}\;\;,
\qe
in which the ``infinitely steep'' mexican-hat form of the potential is evident.

\subsection{The CJT effective potential}

In order to compute thermodynamic
quantities we use the CJT formalism \cite{CJT} which is a self-consistent
resummation prescription to compute the effective potential
for a given theory (see also 
e.g.~Refs.~\cite{Fejos:2009dm,seel,dirk,2pi_thermodynamics_1,2pi_thermodynamics_2,Petropoulos:1998gt,Grahl:2011yk,vanHees:2002bv,AmelinoCamelia:1997dd}).

The first step consists in shifting the fields $\sigma$ and $\alpha$ by their
nonvanishing vacuum expectation values,
\eq
	\sigma\rightarrow\phi+\sigma\;\;,\;\;\alpha\rightarrow\;\alpha_{0}
		+\alpha\;\;.
\qe
As a consequence, the Lagrangian (\ref{lalfa}) takes the form
\eqa
	\mathcal{L}  &=&
\tfrac{1}{2}(\partial_{\mu}\Phi)^{2}
		+\dfrac{N\varepsilon}{8}(\alpha_{0}+\alpha)^{2}
		+\dfrac{i}{2}(\alpha_{0}+\alpha)\left(  \sigma^{2}+\vec{\pi}^{2}+2\sigma
		\phi\ +\phi^{2}-\dfrac{N}{g^{2}}\right)\;\,.
	\label{u1}
\qea
As one can see from Eq.~(\ref{u1}) this shift produces a bilinear mixing term,
$i\alpha\sigma\phi$, which renders the propagator matrix non-diagonal in the
fields $\sigma$ and $\alpha$. There are two possible ways to handle the ensuing
mixing. One is to keep this term and allow for a non-diagonal propagator which
mutually transforms the fields $\sigma$ and $\alpha$
into each other \cite{Cooper:2005vw, Fejos:2009dm}. The
second way, used in this work and based on the study of Ref.~\cite{seel},
consists in performing the shift
\eq
	\alpha\longrightarrow\alpha-4\dfrac{i\phi}{N\varepsilon}\,\sigma\;\;,
	\label{shift}
\qe
which eliminates the mixing term. In Ref.~\cite{seel} it was explicitly
demonstrated up to two-loop order that these two methods to handle the mixing
term are equivalent, i.e.~they yield the same results for the effective
potential and for the gap equations. However, the shift introduced in 
Eq.~(\ref{shift}) considerably simplifies the calculations. From the resulting
Lagrangian,
\eqa
	\mathcal{L} &=& \dfrac{1}{2}\left( \partial _{\mu }\sigma \right) ^{2}+\dfrac{1
		}{2}\left( \partial _{\mu }\pi _{i}\right) ^{2}+\dfrac{1}{2}\left( i\alpha
		_{0}+\dfrac{4\phi ^{2}}{N\varepsilon }\right) \sigma ^{2}+\dfrac{1}{2}\left(
		i\alpha _{0}\right) \pi _{i}^{2}+\dfrac{1}{2}\,\frac{N\varepsilon }{4}
		\,\alpha ^{2}
	\nonumber\\
 	& & +\dfrac{i}{2}\alpha (\sigma ^{2}+\pi _{i}^{2})+\dfrac{2\phi }{N\varepsilon 
		}\,\sigma (\sigma ^{2}+\pi _{i}^{2})+U(\phi ,\alpha _{0})\;\;,
	\label{lag}
\qea
one can immediately deduce the inverse tree-level propagators and the
tree-level masses:
\eqa
	D_{i}^{-1}(K;\phi,\alpha_{0})  &=& K^{2}+m_{i}^{2}\;\;\;,\;\;
		i=\sigma,\;\vec{\pi};\label{di}\\
	m_{\sigma}^{2}  &=& i\alpha_{0}+\dfrac{4\phi^{2}}{N\varepsilon}\;\;\;;
		\;\;\;m_{\pi}^{2}=i\alpha_{0}\;\;;\label{mtr}\\
	D_{\alpha}^{-1}  &=& m_{\alpha}^{2}=\frac{N\varepsilon}{4}\;\;.\label{dtr}
\qea

Within the aforementioned CJT formalism the standard expression for the
effective potential is given by
\eqa
	V_\mathrm{eff} &\equiv& V_\mathrm{eff}(\phi,\alpha_{0};G_{\sigma},G_{\pi},G_{\alpha})=
		\nonumber \\
	& &
		U(\phi,\alpha_{0})+
		\sum_{i=\sigma,\vec{\pi},\alpha}\tfrac{1}{2}\int_{K}[\ln G_{i}^{-1}
	(K)+D_{i}^{-1}(K;\phi,\alpha_{0})G_{i}(K)-1]+V_{2}\;\;,
		\label{infg}
\qea
where $U(\phi,\alpha_{0})$ is the tree-level potential just as in 
Eq.~(\ref{lalfa}), $D_{i}(K;\phi,\alpha_{0})$ are the tree-level propagators,
$G_{i}(K)$ are the full propagators in momentum space and $V_{2}$ contains all
two-particle-irreducible (2PI) self-interaction terms. The full propagators and the
condensates are evaluated by using the stationary conditions for the effective
potential:
\eq
	\dfrac{\delta V_\mathrm{eff}}{\delta\phi}=0\;\;,\ \ \dfrac{\delta V_\mathrm{eff}}
		{\delta\alpha_{0}}=0\;\;,\;\;\dfrac{\delta V_\mathrm{eff}}{\delta G_{i}
		(K)}=0\;\;;\ \ \ i=\sigma,\vec{\pi},\alpha\;\;.
	\label{stateq}
\qe
For the full propagators one obtains
\eq
	G_{i}^{-1}(K)=K^{2}+M_{i}^{2}\;\;\;;\;\;\;M_{i}^{2}\ =m_{i}^{2}+\Sigma_{i}\;\;,
\qe
where $\Sigma_{i}$ denotes the self-energy
\eq
	\Sigma_{i}=2\dfrac{\delta V_{2}}{\delta G_{i}(K)}\;\;.
\qe

\subsection{One-loop approximation}
Restricting to one-loop order, the 2PI contribution to the effective potential
is equal to zero, $V_{2}\equiv0$.
Thus, the effective potential is given by:
\eq
	V_\mathrm{eff}=\dfrac{i}{2}\alpha_{0}\left(  \phi^{2}-\dfrac{N}{g^{2}}\right)
		+\dfrac{N\varepsilon}{8}\alpha_{0}^{2}\
		+\dfrac{1}{2}\sum_{i=\sigma,\vec{\pi},\alpha}\int_{K}[\ln G_{i}^{-1}
		(K)+D_{i}^{-1}(K;\phi,\alpha_{0})G_{i}(K)-1]\;\;.
\qe
Using the stationary conditions (\ref{stateq}) one derives the following
equations for the two condensates:
\eqa
	0 &=& \phi\left(  i\alpha_{0}+\dfrac{4}{N\varepsilon}\int_{K}G_{\sigma}(K)\;\;\right)  \;\;; \\
	i\alpha_{0} &=& \dfrac{2}{N\varepsilon}\left[  \phi^{2}-\frac{N}{g^{2}}
		+\int_{K}G_{\sigma}(K)+(N-1)\int_{k}G_{\pi}(K)\right]  \;\;,\label{a1}
\qea
and for the full propagators:
\eqa
	G_{i}^{-1}(K) &=& K^{2}+M_{i}^{2}\ =D_{i}^{-1}(K;\phi,\alpha_{0})\;\;;\\
	M_{i}^{2}\    &=& m_{i}^{2}+\Sigma_{i}=m_{i}^{2}\;\;.
\qea
In fact, since $V_{2}=0$, the self-energy $\Sigma_{i}$ vanishes in the
one-loop approximation. Substituting the right hand side of
Eq.~(\ref{a1}) for $i\alpha_{0}$ gives the following equations for the physical condensate
$\phi$ and for the masses:
\eqa
	0 &=& \phi\left(  M_{\pi}^{2}+\dfrac{4}{N\varepsilon}\int_{K}G_{\sigma
		}(K)\right)  \;\;;\;\label{gap1}\\
	M_{\sigma}^{2} &=& M_{\pi}^{2}+\dfrac{4\phi^{2}}{N\varepsilon}\;\;;\label{gap3}\\
	M_{\pi}^{2} &=& \frac{2}{N\varepsilon}\left[  \phi^{2}-\frac{N}{g^{2}}
		+\int_{K}G_{\sigma}(K)+(N-1)\int_{K}G_{\pi}(K)\right]  \;\;.\label{gap4}
\qea
We now study separately the cases $\phi\neq0$ and $\phi=0$ and show that, in
the present two-dimensional model, the latter holds.

\begin{enumerate}
\item[(i)]{ We assume $\phi\neq0$ (corresponding to spontaneous symmetry breaking).
When taking the limit $\varepsilon\rightarrow0^{+}$ the degree of
freedom denoted by $\sigma$ becomes frozen due to an infinitely heavy mass,
see Eq.~(\ref{gap3}):
\eq
	\lim_{\varepsilon\rightarrow0^{+}}M_{\sigma}^{2}=\lim_{\varepsilon
		\rightarrow0^{+}}\dfrac{4\phi^{2}}{N\varepsilon}=\infty\;\;,
\qe
from which it follows that
\eq
	\lim_{\varepsilon\rightarrow0^{+}}\dfrac{1}{\varepsilon}\int_{K}G_{\sigma}(K)=0\;\;.
\qe
Moreover, from Eq.~(\ref{gap1}) we obtain that $M_{\pi}=0$.
Then, the gap equation (\ref{gap4}) becomes:
\eq
	\phi^{2} = \frac{N}{g^{2}}-(N-1)\int_{K}G_{\pi}(K) =
		\frac{N}{g^{2}}-(N-1)\int_{0}^{\infty}\dfrac{\de k}{\pi}
		\dfrac{1}{k\left[\exp\left(  k/T\right)  -1\right]  }\;\;.
	\label{case1}
\qe
There is no solution to this equation since the integral in the right-hand side
is divergent, whereas $\phi^{2}-N/g^{2}$ is finite. Thus, the case $\phi\neq0$
leads to inconsistencies and cannot hold. This means that in two
dimensions there is no spontaneous symmetry breaking of the global $O(N)$
symmetry of the nonlinear $O(N)$ model,
in agreement with the Mermin-Wagner-Coleman theorem mentioned in the Introduction.
Note that in four dimensions this would have not been the
case: a condensation of $\phi$ would take place signalling an explicit breaking of
chiral symmetry, see for instance Refs.~\cite{seel,dirk,andersen}.
}

\item[(ii)] {We set $\phi=0$ (no spontaneous symmetry breaking). In this case 
Eq.~(\ref{gap1}) is trivially fulfilled. Eq.~(\ref{gap3}) implies that the masses
of $\sigma$ and $\pi$ become degenerate,
\eq
	M_{\sigma}^{2}=M_{\pi}^{2}\equiv M^{2}\;\;,
\qe
so that
\eq
	G_\pi(K)=G_\sigma(K)\equiv G(K)\;\;.
\qe
Then, from Eq.~(\ref{gap4}) we find
\eq
	\frac{N}{g^{2}}=N\int_{K}G(K)=N\int_{0}^{\infty}\dfrac{\de k}{\pi}~\dfrac{n\left[
		\omega_{k}\left(  M\right)  \right]  }{\omega_{k}\left(  M\right)  }\;\;,
\label{gap}
\qe
where
\eq
\omega_{k}\left(  M\right)  =\sqrt{k^{2}+M^{2}}
\qe
and
\eq
	n\left[  \omega_{k}\left(  M\right)  \right]  =
		\left\{  \exp\left[  \omega_{k}\left(  M\right)  /T\right]  -1\right\}  ^{-1}\;\;.
\qe
From the previous equation one can determine the behaviour of the function
$M(T)$, see later. The case $\phi=0$ does not lead to any inconsistencies and
is the realised one. It should be stressed that this solution is highly
nontrivial: instead of two massless degrees of freedom one has in the vacuum three
equally massive particles.

The basic quantity for a thermodynamic study of the system is the pressure,
which is, up to a sign, identical to the minimum of the effective potential,
\begin{displaymath}
	p=-V_\mathrm{eff}^\mathrm{min}\;\;.
\end{displaymath}
From the pressure we can compute the energy density $\epsilon$ and the trace
anomaly $\theta$ using the first principle of thermodynamics:
\begin{eqnarray*}
	\epsilon &=& T\frac{dp}{dT}-p\;\;;\\
	\theta &=& \epsilon-p\;\;.
\end{eqnarray*}
}
\end{enumerate}

\subsection{Regularisation}
The nonlinear $O(N)$ model is infrared-finite, since the mass is
nonvanishing. Thus, there are only ultraviolet divergences, which we
regularise via an ultraviolet cutoff $\Lambda$ \cite{novikov, warringa1, warringa2}:
\eqa
	\int_{K}G(K)&=& \int_{0}^{\Lambda}\dfrac{\de k}{\pi}~\dfrac{n\left[  \omega_{k}
		\left(  M\right)  \right]  }{\omega_{k}\left(  M\right)  }+\frac{1}{4\pi}
		\ln\frac{\Lambda^{2}}{M^{2}}\;\;; \\
	\int_{K}\ln G^{-1}(K)  &=& -2\int_{0}^{\Lambda}\frac{\de k~k^{2}}{\pi}~
		\dfrac{n\left[  \omega_{k}\left(  M\right)  \right]  }{\omega_{k}\left(
		M\right)  } +\frac{M^{2}}{4\pi}\left(  1+\ln\frac{\Lambda^{2}}{M^{2}}\right)  \;\;,
	\label{infg1}
\qea
where we subtracted divergent $M$- and $T$-independent terms. In the previous
equations the quantity $M=M(T)$ denotes the temperature-dependent mass. At $T=0$ one has
$M(T=0)=m$, with $m$ being the only dimensionful parameter of the model (emerging
through dimensional transmutation) --- see the next Subsection for further details.

Using equations (\ref{infg}) and (\ref{infg1}) we obtain the regularised
effective potential and the regularised pressure:
\eq
	p_\mathrm{reg} \equiv p_\mathrm{reg}(T)  =\dfrac{NM^{2}}{2g^{2}}+N\int_{0}^{\Lambda}\frac{\de k~k^{2}}{\pi}
		~\dfrac{n\left[  \omega_{k}\left(  M\right)  \right]  }{\omega_{k}\left(
		M\right)  } -\frac{NM^{2}}{8\pi}\left(  1+\ln\frac{\Lambda^{2}}{M^{2}}\right)\;\;,
	\label{preg}
\qe
where the regularised mass $M=M(T)$ is determined from
\eq
	\frac{N}{g^{2}}=N\int_{0}^{\Lambda}\frac{\de k}{\pi}
		~\dfrac{n\left[ \omega_{k}\left(  M\right)  \right]  }{\omega_{k}
		\left(  M\right)  }+\frac{N}{4\pi}\ln\frac{\Lambda^{2}}{M^{2}}\;\;.
\qe
Note that at $T=0$ we obtain the following relation:
\eq
	\frac{N}{g^{2}}=\frac{N}{4\pi}\ln\frac{\Lambda^{2}}{m^{2}}\;\;.
	\label{rel}
\qe

\subsection{Renormalisation}
After regularisation we can perform the renormalisation. First, we compute the
renormalised coupling constant as it was defined in \cite{novikov}:
\eqa
	\frac{1}{g_\mathrm{ren}^{2}} &=& \frac{2}{N}\frac{\de p_\mathrm{reg}(T=0)}{\de m^{2}}\Big|_{m^{2}=\mu^{2}}
		\nonumber \\
	&=& \frac{2}{N}\frac{\de}{\de m^{2}}\left[  N\dfrac{m^{2}}{2g^{2}}-N\frac{m^{2}}{8\pi}
		\left(  1+\ln\frac{\Lambda^{2}}{m^{2}}\right)
		\right]_{m^{2}=\mu^{2}}=\frac{1}{g^{2}}-\frac{1}{4\pi}\ln\frac{\Lambda^{2}}{\mu^{2}}\;\;.
	\label{gren}
\qea
Inserting this result in Eq.~(\ref{preg}) and subtracting the zero-temperature
contribution $-Nm^{2}/8\pi$ we get the renormalised pressure in the
limit $\Lambda\rightarrow\infty$:
\eqa
	p_\mathrm{ren}  &=& \dfrac{NM^{2}}{2g^{2}}+N\int_{0}^{\infty}\frac{\de k~k^{2}}{\pi}
		~\dfrac{n\left[  \omega_{k}\left(  M\right)  \right]  }{\omega_{k}\left(M\right)  }
		-\frac{NM^{2}}{8\pi}\left(  1+\ln\frac{\Lambda^{2}}{M^{2}}\right)+\frac{Nm^{2}}{8\pi} = \nonumber\\
	&=& \dfrac{NM^{2}}{2g_\mathrm{ren}^{2}}+N\int_{0}^{\infty}\frac{\de k~k^{2}}{\pi}
		~\dfrac{n\left[  \omega_{k}\left(  M\right)  \right]  }{\omega_{k}\left(M\right)  }
		-\frac{NM^{2}}{8\pi}\left(  1+\ln\frac{\mu^{2}}{M^{2}}\right)
		+\frac{Nm^{2}}{8\pi}\;\;, \label{pren}
\qea
where we have used Eq.~(\ref{rel}). The temperature-dependent mass $M$ is
determined from the renormalised gap equation:
\eqa
	\frac{1}{g^{2}}  &=& \frac{1}{g_\mathrm{ren}^{2}}+\frac{1}{4\pi}\ln\frac{\Lambda^{2}}{\mu^{2}}
		=\int_{0}^{\infty}\frac{\de k}{\pi}~\dfrac{n\left[  \omega_{k}\left(  M\right)\right]  }{\omega_{k}
		\left(  M\right)  }+\frac{1}{4\pi}\ln\frac{\Lambda^{2}}{M^{2}}\;\;;\nonumber\\
	\frac{1}{g_\mathrm{ren}^{2}} &=& \int_{0}^{\infty}\frac{\de k}{\pi}~\dfrac{n\left[\omega_{k}\left(  M\right)  \right]  }{\omega_{k}
		\left(  M\right)  }+\frac{1}{4\pi}\ln\frac{\mu^{2}}{M^{2}}\;\;. \label{rencond}
\qea
Solving Eq.~(\ref{rencond}) at $T=0$ one can show that the model exhibits
dimensional transmutation, meaning that there is a nonvanishing mass parameter
$m$ in the vacuum which is generated due to the renormalisation of quantum
corrections:
\eq
	M^{2}(T=0)=m^{2}=\mu^{2}\exp\left(-\frac{4\pi}{g_\mathrm{ren}^{2}}\right)\;\;.
\qe
Notice that $m$ depends non-analytically on the coupling constant for
$g\rightarrow0$. Thus, the mass would vanish if naive perturbation theory were applied.

In order to determine the asymptotic behaviour of the theory we compute the
beta-function of the running coupling:
\eqa
	\mu\frac{\de g_\mathrm{ren}^{2}}{\de\mu} &=& \mu\frac{\de}{\de\mu}\left[  g^{2}\left(
		1-\frac{g^{2}}{4\pi}\ln\frac{\Lambda^{2}}{\mu^{2}}\right)  ^{-1}\right] = \nonumber \\
	&=& -\frac{1}{2\pi}\left[  g^{2}\left(  1-\frac{g^{2}}{4\pi}\ln\frac
		{\Lambda^{2}}{\mu^{2}}\right)  ^{-1}\right]  ^{2} =-\frac{g_\mathrm{ren}^{4}}{2\pi}<0\;\;.
\qea
We have shown that the coupling constant decreases with increasing $\mu$ and
thus the theory is asymptotically free.

\subsection{Two-loop approximation}

The Lagrangian (\ref{lag}) contains only three-point interaction terms and no
four-point vertices. Therefore to two-loop order we can only construct sunset
diagrams but no double-bubbles. Besides, since there is no spontaneous
symmetry breaking in two dimensions, i.e.~$\phi=0$ and $M_{\sigma}^{2}=M_{\pi
}^{2}=M^{2},$ only the term $i\alpha(\sigma^{2}+\pi_{i}^{2})/2$ contributes to
the 2PI part of the effective potential via a sunset diagram:
\eq
V_{2}=\frac{N}{4}\int_{K}\int_{P}G_{\alpha}(K+P)G(K)G(P)\;\;.
\qe
Taking the sunset diagram into account the effective potential to two-loop
order is:
\eqa
	V_\mathrm{eff}(\alpha_{0},G,G_{\alpha}) &=& N\dfrac{i\alpha_{0}}{2g^{2}}
		+\dfrac{N\varepsilon}{8}\alpha_{0}^{2}
	\label{v2}
	\\
		& & ~~ +\dfrac{1}{2}\sum_{i=\sigma,\vec{\pi},\alpha}\int_{K}[\ln G_{i}^{-1}
		(K)+D_{i}^{-1}(K;\alpha_{0})G_{i}(K)-1]\;\;+V_{2}\;\;.
		\nonumber
\qea
The corresponding equations for the condensate of the auxiliary field and for
the full propagators are:
\eqa
	i\alpha_{0}  &=& \dfrac{2}{N\varepsilon}\left(  -\frac{1}{g^{2}}+\int_{K}G(K)\right)  \;\;;\label{alpha}\\
	G^{-1}(K)  &=& D^{-1}(K)+\Sigma(K)\;\;;\\
	G_{\alpha}^{-1}(K)  &=& D_{\alpha}^{-1}(K)+\Sigma_{\alpha}(K)\;\;,
\qea
where the self-energies for $\alpha$ and for the $N$ scalar fields read
\eqa
	\Sigma(K)  &=& \frac{2}{N}\dfrac{\delta V_{2}}{\delta G(K)}=\int_{P}G(P)G_{\alpha}(K-P)\;\;;\\
	\Sigma_{\alpha}(K)  &=& 2\dfrac{\delta V_{2}}{\delta G_{\alpha}(K)}=\frac{N}{2}\int_{P}G(P)G(K-P)\;\;.
\qea
The temperature-dependent mass of the scalar fields is determined by solving
\eq
	M^{2}  =m^{2}+\operatorname{Re}\Sigma=i\alpha_{0}+\operatorname{Re}\int_{P}G(P)G_{\alpha}(K-P)\;\;,
	\label{mpi2}
\qe
where we have used $m^{2}=i\alpha_{0},$ see Eq.~(\ref{mtr}). Besides, using
Eq.~(\ref{dtr}) we obtain the following relation for the full propagator of
the auxiliary field:
\eq
	G_{\alpha}^{-1}(K)=\frac{N\varepsilon}{4}+\frac{N}{2}\int_{P}G(P)G(K-P)\;\;.
	\label{malpha}
\qe
Next, we perform the following three steps to simplify these results:

\begin{enumerate}
\item[(i)] {Out of Eqs.~(\ref{alpha}), (\ref{mpi2}), and (\ref{malpha}) we obtain
\eqa
	-\frac{1}{g^{2}}+\int_{K}G(K) &=& 2\varepsilon\left[  M^{2}-\operatorname{Re}
		\int_{P}G(P)G_{\alpha}(K-P)\right] \nonumber \\
	&=& 2\varepsilon\left[  M^{2}-\operatorname{Re}\int_{P}\frac{G(P)}{ \frac{N\varepsilon}{4}
		+\frac{N}{2}\int_{L}G(L)G(K-P-L)}\right]  \;\;.
\qea
In the nonlinear limit $\varepsilon$ $\rightarrow0^{+}$ this expression
reduces to
\eqa
	\frac{1}{g^{2}}  &=& \int_{K}G(K)
		=\int_{K}\frac{1}{K^{2}+i\alpha_{0}+\int_{P}G(P)G_{\alpha}(K-P)} \nonumber \\
	&=& ~~ \int_{K}\frac{1}{K^{2}+M^{2}+\operatorname{Im}\int_{P}G(P)G_{\alpha}(K-P)}\;\;,
\qea
where the real part of the sunset has been reabsorbed in the definition of the
mass, see Eq.~(\ref{mpi2}). Moreover, since all three particles have the same
mass and decays are strongly suppressed, we neglect the imaginary part of the
self-energy:
\eq
	\operatorname{Im}\left[  \int_{P}G(P)G_{\alpha}(K-P)\right]  =0\;\;.
\qe
Then, we obtain at two-loop order the same gap equation as in the one-loop approximation:
\eq
	\frac{1}{g^{2}}=\int_{K}G(K)\;\;; \label{gap2}
\qe
this result is consistent with the observation that the
next-to-leading-order effective potential is minimised by the leading-order mass
\cite{warringa1,warringa2, root}.
}
\item[(ii)] {Furthermore, using Eqs.~(\ref{dtr}) and (\ref{malpha}) one can show that
in the nonlinear limit the two integrals $\int_{K}D_{\alpha}^{-1}G_{\alpha}(K)-1$
and $\int_{K}\int_{P}G_{\alpha}(K+P)G(K)G(P)$, involving the auxiliary
field appearing in the effective potential (\ref{v2}), reduce to constant
terms which are irrelevant for the thermodynamics:
\eq
	\int_{K}D_{\alpha}^{-1}G_{\alpha}(K)-1  =\frac{\frac{N\varepsilon}{4}}{\frac{N\varepsilon}{4}
		+\frac{N}{2}\int_{P}G(P)G(K-P)}-1\nonumber\\
		\underset{\varepsilon\rightarrow0^{+}}{=}-1\;\;.
	\label{c1}
\qe
Similarly, we get (at the minimum):
\eqa
	V_{2}^\mathrm{min}  &=& \frac{N}{4}\int_{K}\int_{P}G_{\alpha}(K)G(P)G(K+P)\nonumber\\
	&=& \frac{N}{4}\int_{K}\frac{1}{\frac{N\varepsilon}{4}+\frac{N}{2}\int_{L}G(L)G(K-L)}\int_{P}G(P)G(K+P)\nonumber\\
	&\underset{\varepsilon\rightarrow0^{+}}{=}& \frac{N}{4}\int_{K}\frac{1}{\frac{N}{2}\int_{L}G(L)G(K-L)}\int_{P}G(P)G(K+P)\nonumber\\
	&\underset{L=-Q}{=}& \frac{N}{4}\int_{K}\frac{1}{\frac{N}{2}\int_{-Q}G(Q)G(K+Q)}\int_{P}G(P)G(K+P)\nonumber\\
	&=& \frac{N}{4}\int_{K}\frac{1}{\frac{N}{2}\int_{Q}G(Q)G(K+Q)}\int_{P}G(P)G(K+P)=\mathrm{const.}
\qea
}
\item[(iii)] {Finally, we can rewrite the term $\int_{K}\left[  D^{-1}(K;\alpha
_{0})G(K)-1\right]  $ as follows:
\eqa
	\int_{K}\left[  D^{-1}(K;\alpha_{0})G(K)-1\right]
	&=& \int_{K}\left[  \frac{K^{2}+m^{2}}{K^{2}+M^{2}}-1\right] = \label{c3} \\
	&=& \int_{K}\frac{m^{2}-M^{2}}{K^{2}+M^{2}}
	=\left(  m^{2}-M^{2}\right)  \int_{K}G(K)
	=\frac{\left(  m^{2}-M^{2}\right)  }{g^{2}}\;\;,\nonumber
\qea
where we used the gap equation (\ref{gap2}). Besides, we set $D^{-1}
(K;\alpha_{0})=K^{2}+i\alpha_{0}=K^{2}+m^{2}.$
}
\end{enumerate}

Using the results obtained in (i), (ii) and (iii) the effective potential to
two-loop order simplifies as follows:
\eq
V_\mathrm{eff}^\mathrm{min}=-\frac{NM^{2}}{2g^{2}}+\dfrac{N}{2}\int_{K}\ln G^{-1}(K)+\dfrac{1}
{2}\int_{K}\ln G_{\alpha}^{-1}(K)\;\;. \label{veff2}
\qe

At $T=0$ the effective potential can be regularised analytically. Introducing
an ultraviolet momentum cutoff and dropping $m$-independent divergences one
obtains \cite{novikov, warringa1, warringa2}:
\eqa
	V_\mathrm{eff}(T=0)  &=& -\frac{Nm^{2}}{2g^{2}}+\frac{Nm^{2}}{8\pi}\left(1+\ln\frac{\Lambda^{2}}{m^{2}}\right) \\
	& & ~~ +\frac{1}{8\pi}\left[  \left(  \Lambda^{2}+2m^{2}\right)  \ln\ln
		\frac{\Lambda^{2}}{m^{2}}-m^{2}\operatorname{li}\frac{\Lambda^{2}}{m^{2}
		}\right]
		+\frac{m^{2}}{4\pi}\left[  \gamma_{E}-1-\ln\frac{\Lambda^{2}}{4m^{2}}\right]
		\;\;,\nonumber
\qea
with $\mathrm{li}(x)$ denoting the logarithmic integral function.

However, at finite temperature numerical methods must be employed in order to
calculate thermodynamical quantities. As was computed in \cite{warringa1,
warringa2}, the final result for the renormalised pressure to two-loop order
at nonzero $T$ is given by:
\eqa
	p_\mathrm{ren}^\mathrm{(2-loop)} &=& \frac{N-2}{8\pi}\left(  m_{\phi}^{2}-M_{\phi}^{2}\right)
		+N\int_{0}^{\infty}\frac{\de k k^{2}}{\pi}~\dfrac{n\left[  \omega_{k}\left(
		M_{\phi}\right)  \right]  }{\omega_{k}\left(  M_{\phi}\right)  }  \\
	\label{pren2loop}
	& & ~~ +\frac{NM_{\phi}^{2}}{2}\int_{0}^{\infty}\frac{\de k}{\pi}~\dfrac{n\left[
		\omega_{k}\left(  M_{\phi}\right)  \right]  }{\omega_{k}\left(  M_{\phi
		}\right)  }
		+\frac{1}{2}\left[  M_{\phi}^{2}\frac{\de \left(  F_{1}+F_{2}\right)  }{\de M_{\phi}^{2}}-F_{1}-F_{2}\right]  \;\;. \nonumber
\qea
Here $F_{1}$ is calculated using the Abel-Plana formula
\eq
	F_{1}=T\sum_{k_{\tau}=2\pi nT}\int_{-\infty}^{+\infty}\frac{\de k}{2\pi}
		\ln G_{\alpha}^{-1}(K)-\int_{-\infty}^{+\infty}
		\frac{\de k_\tau}{\left(  2\pi\right)} \int_{-\infty}^{+\infty} \frac{\de k}{\left(  2\pi\right)}\ln G_{\alpha}^{-1}(K)\;\;,
\qe
and $F_{2}$ takes the form
\eq
F_{2}=\mathcal{P}\int_{K}\ln\left[  \frac{\widetilde{G}_{\alpha}^{-1}(K)}
{\ln\left(  K^{2}/\widetilde{M}^{2}\right)  }\right]  -\frac{M^{2}}{2\pi}\ln
\ln\left(  \frac{\Lambda^{2}}{\widetilde{M}^{2}}\right)  \;\;,
\qe
where $\mathcal{P}$ is indicates the principal value of the integral, and
\eqa
	G_{\alpha}^{-1}(K)  &\equiv& \frac{1}{4\pi}\frac{\widetilde{G}_{\alpha}^{-1}
		(K)}{\sqrt{K^{2}(K^{2}+4M^{2})}}\;\;;\\
	\widetilde{M}^{2}  &=& M^{2}\exp\left[  -4\int_{0}^{\infty}
		\dfrac{\de k}{\omega_{k}~(M_{\phi})}~\dfrac{1}{\exp\left[  \omega_{k}(M_{\phi})/T\right]-1}\right] \;\;.
\qea
Besides, $M_{\phi}$ is obtained by solving the equations
\eqa
	\frac{4\pi}{g_\mathrm{ren}^{2}}  &=& \left(  1-\frac{2}{N}\right)  \ln\frac{\mu^{2}
		}{\overline{M}_{\phi}^{2}}
		+\frac{8}{N}\int_{0}^{\infty}\frac{\de k}{\pi}~\dfrac{1}{\omega_{k}(M_{\phi}
		)}~\dfrac{1}{\exp\left[  \omega_{k}(M_{\phi})/T\right]  -1}\nonumber\\
		& & ~~ +\frac{2}{N}\left[  \ln4+2\pi\frac{\de \left(  F_{1}+F_{2}\right)  }{\de M_{\phi}^{2}}\right]  \;\;;\\
	\overline{M}_{\phi}^{2}  &=& M_{\phi}^{2}\exp\left[  -4\int_{0}^{\infty}
		\dfrac{\de k}{\omega_{k}}~\dfrac{1}{\exp\left[  \omega_{k}/T\right]  -1}\right]\;\;.
\qea
These expressions
will be used in Sec.~\ref{sec:comparison} when comparing to the lattice results.
Note that our results for the pressure to one- and two-loop order turn out to be
\emph{identical} to the ones of Ref.~\cite{warringa1} in the $1/N$ expansion.
This is a quite surprising result, which we discuss in more detail. Namely, in
general the CJT formalism and the $1/N$ expansion are two different
resummation prescriptions. The $1/N$ expansion corresponds to the large-$N$
limit, whereas the CJT formalism is a loop expansion (moreover, the CJT
formalism is applicable to systems out of equilibrium, which is not the case
for the $1/N$ expansion).

Considering the case of 1+3 dimensions, one can show that in general these
two resummation prescriptions give different results: to LO in the $1/N$
expansion there are $N$ degenerate (and, in the chiral limit, massless) pions
and no massive sigma field is present, since this excitation only starts to
propagate at NLO, see Ref.~\cite{2pi_thermodynamics_1}.
Conversely, in the CJT formalism in the one-loop
approximation there are $N-1$ massive pions and one massive sigma field, see
Ref.~\cite{seel}: this example demonstrates that the two methods are different
(note that, if we perform the large $N$-limit \emph{within} the one-loop order
of the CJT method, there are $N$ pions left. Thus, only if the large $N$
limit is additionally applied does the one loop approximation of the
CJT formalism give the same results as the LO in the $1/N$ expansion).

The fact that in two dimensions the results are equivalent, as shown in this
work, is due to the absence of spontaneous symmetry breaking. Since the
condensate is zero, the sigma and the pion become degenerate. This is equivalent
to the large-$N$ limit with $N$ degenerate particles, which explains the
agreement of our results with the LO approximation in $1/N$.

For the agreement at two loop order in 1+1 dimensions we were not able to
get an analogous explanation: the two-loop result for the effective
potential we obtain in Eq.~(\ref{v2}) has a different expression from the one
obtained in the $1/N$ expansion at NLO, which is equivalent to Eq.~(\ref{veff2}).
Even after taking into account the absence of spontaneous symmetry breaking,
one can not see directly the exact differences between the CJT formalism and
the $1/N$ expansion. Only after taking the nonlinear limit, which is related
to an infinitely large coupling constant, is it possible to perform some analytical
simplifications, which allow us to rewrite the effective potential in such a
way that it has the same form as the one in the $1/N$ expansion, Eq.~(\ref{veff2}).
Moreover, what makes the comparison more difficult is the fact that we handle
the bilinear propagator in a different way than the authors of Refs.~\cite{warringa1,warringa2}.
We eliminate the mixing term by a shift of the auxiliary field, while the
authors of Refs.~\cite{warringa1,warringa2}
keep the mixing term and work with non-diagonal propagators.

Summarising, one can state that the exact determination of the differences
between the CJT formalism and the $1/N$ expansion is a nontrivial task, which
requires a detailed analytical study order by order. The two methods in
general differ, since the leading orders in 1+3 dimensions are different;
the fact that they coincide to leading order in 1+1 dimensions is also
due to the peculiar dimensionality. The reason why they also
coincide at two-loop order could not be clearly isolated, although we suspect
that the lower dimensionality and the nonlinear limit play an important r\^ole.
A further open question is whether this equivalence would persist at higher orders.
While an explicit calculation is prohibitive, progress in this direction can
be made if a formal way to connect the different methods is found.

\section{Lattice simulation}
\label{sec:montecarlo}

In this Section we present the detailed evaluation of thermodynamical
quantities of the $O(3)$ model by numerical lattice calculation. 
Some of the model's thermodynamic properties were studied in \cite{Spiegel:1996fr}
in the context of the quest for a ``perfect action'', free of discretisation systematics
(see \cite{perfect_action_1,perfect_action_2}); a direct comparison to those findings seems however not to have
an immediate significance, since there the fixed ratio $N_x/N_t=3$ is maintained,
while we choose to keep it well above $10$ in all our simulations in order to
suppress corrections to the infinite-volume limit: this difference
is important enough to alter the high-temperature asymptotic behaviour
of $p/T^2$, thereby invalidating any numerical comparison.
Other than Ref.~\cite{Spiegel:1996fr}, to our knowledge no dedicated study of the thermodynamics
has been performed on this model.

The
Euclidean, discretised action under study takes the form of a Heisenberg
model, i.e.~an $O(3)$ spin model with nearest-neighbour interaction:
\eq
S=\beta\sum_{\langle i,j\rangle}\left(  1-\vec{s}_{i}\cdot\vec{s}_{j}\right)
\;, \label{eq:s_latt}
\qe
where the sum runs over all bonds of a 2-dimensional lattice and the $\vec
{s}_{i}$ are 3-dimensional unit vectors in internal space. The system at
finite temperature $T$ is realised by making the timelike extent of the
lattice finite and consisting of $N_{t}$ sites, with periodic boundary
conditions in that direction: denoting with $a$ the lattice spacing, we have
$aN_{t}=1/T$. The identification of the above action with the continuum one
defining the partition function Eq.~(\ref{z}) is completed by setting $\beta=N/g^2$.

In a practical Monte Carlo simulation, moreover, the space extent of the
lattice, $N_{x}$, is also necessarily finite; as long as $N_{x}\gg N_{t}$,
however, this will hardly affect the temperature setting. In order to minimise
the finite-size corrections, we will assume periodic boundary conditions in
the spacelike direction throughout all of the following.

Our goal is to extract the thermodynamical behaviour of the action
Eq.~(\ref{eq:s_latt}). In the next parts, the lattice methods will be
illustrated along with the determination of the physical scale; subsequently,
we present the numerical results that will later be compared to the predictions of the
previous Section. A discussion of the algorithms implemented, their efficiency
and autocorrelation is deferred to Appendix \ref{app:algorithms}.

\subsection{Thermodynamical observables}

For the evaluation of pressure $p$, trace anomaly $\theta$ and energy density
$\epsilon$, we adapt the ``integral method'' \cite{boyd} to our
(1+1)-dimensional model. We start from the continuum expression for the
(nonphysical) pressure,
\eq
	p_{*}(T)= T \frac{\partial\ln\mathcal{Z}}{\partial V}=\frac{T}{V}
		\ln\mathcal{Z}
	\label{eq:pressure_start}
\qe
(the second equality stems from the assumption of spatial isotropy, that
always holds in the thermodynamic limit), where $\mathcal{Z}$ is the partition
function associated to $S$ (it is understood that it refers to a system with
timelike extent $N_{t}$):
\eq
\mathcal{Z} \propto\Bigg( \prod_{\ell} \int_{S^{2}} \de\vec{s}_{\ell} \Bigg) \,
\prod_{\langle i,j \rangle} e^{\beta\vec{s}_{i}\cdot\vec{s}_{j}}\;.
\qe
Since the one-dimensional spatial volume is given by $V=aN_{x}$, we have
\eq
a^{2} p_{*}(T) = \frac{1}{N_{x}N_{t}} \ln\mathcal{Z}\;;
\qe
we then write $\ln\mathcal{Z}$ as an integral in order to express it in terms
of measurable quantities,
\eq
a^{2} p_{*}(T) = \int_{\beta_0}^{\beta}\de\beta^{\prime}\frac{1}{N_{x}N_{t}}
\frac{\partial\ln\mathcal{Z}}{\partial\beta}\Big|_{\beta^{\prime}}\;;
\qe
the integrand, exploiting translation invariance, has now become the
expectation value of the energy per site,
\eq
\frac{1}{N_{x}N_{t}}\frac{\partial\ln\mathcal{Z}}{\partial\beta} =
\Big\langle \ell_{x} + \ell_{t} \Big\rangle\;,
\qe
with the shorthand $\ell_{e} = \vec{s}_{i}\cdot\vec{s}_{i+\hat{e}}$. The last
step is an additive renormalisation, needed to make the pressure finite: the
standard choice is \mbox{$p(T) = p_*(T)-p_*(0)$}, which translates into the
subtraction of the same integrand evaluated in a zero-temperature system (that
is, $N_{t}=\infty$): making coupling and time extent explicit in the average
values, we have then
\eq
\frac{p(T)}{T^{2}}=N_{t}^{2}\int_{0}^{\beta}\left(  \left\langle \ell_{x} +
\ell_{t} \right\rangle _{\beta^{\prime},N_{t}} -2\left\langle \ell
\right\rangle _{\beta^{\prime},\infty} \right)  d\beta^{\prime}\;.
\label{eq:presint}
\qe
In the above we fix the integration to start at $\beta_{0}=0$: on the
practical side, this value can be directly simulated with no particular
troubles.\footnote{It is in fact trivial and corresponds to generating all
$\vec{s}_{i}$ uniformly on the unit $2$-sphere.} Note also that $T=0$ will be
numerically approximated with a large enough square system, that is
$N_{x}=N_{t}$.

Likewise, for the trace anomaly, starting from the continuum expression
\eq
\theta=T^{3}\frac{\partial}{\partial T}\left(  \frac{p}{T^{2}}\right)  \;,
\qe
one derives
\eq
\frac{\theta(T)}{T^{2}}=N_{t}^{2}\left(  \frac{\partial\beta}{\partial\ln
T}\right)  \left(  \left\langle \ell_{x}+\ell_{t}\right\rangle _{\beta,N_{t}
}-2\left\langle \ell\right\rangle _{\beta,\infty}\right)  \;.
\label{eq:tracanom}
\qe
This does not require integration over the coupling, but the knowledge of the
beta-function is needed for the prefactor
\eq
\frac{\partial\beta}{\partial\ln T}=-a\frac{\partial\beta}{\partial a}\;.
\label{eq:debeta_delogt}
\qe
The above will be extracted nonperturbatively, measuring the running of the
coupling with the scale and using a suitable parametrisation of the resulting data.

Finally, we also consider the energy density
\eq
\epsilon=\frac{T^{2}}{V}~\frac{\partial\ln\mathcal{Z}}{\partial T}=\theta+p\;.
\label{eq:en_density}
\qe

\subsection{Scale setting}
The running of the scale $a(\beta)$ is needed to convert the temperature in
physical units, as well as for the calculation of $\theta$ and $\epsilon$. We
evaluate it by computing, in a zero-temperature system and over a wide range
of couplings, the spin-spin correlation function
\eq
C(an)=\langle\vec{s}_{i}\cdot\vec{s}_{i+n\hat{e}}\rangle\;, \label{eq:2ptfct}
\qe
which then, to obtain the zero-temperature mass $m$, is fitted to the
expected functional form for a two-dimensional scalar field, valid for large $R$:
\eq
	C(R) \simeq c_{0}\left(  \frac{e^{-mR}}{\sqrt{mR}}+ \frac
		{e^{-m(aN_{x}-R)}}{\sqrt{m(aN_{x}-R)}}\right)  +c_{1}\;,
	\label{eq:correlat}
\qe
where we consider also the ``mirror term'' to account for the periodic
boundaries. Distances on the lattice are in units of $a$, therefore the masses
we measure are the adimensional quantities $am$; once we fix the
corresponding physical value we have the function $a(\beta)$ in units of
$1/m$. The functional form Eq.~(\ref{eq:correlat}), furthermore, is valid
for large enough distances: in practice, we will fit the measurements to the
function in a variable range $R_\mathrm{min} \leq R \leq R_\mathrm{max}$, looking for a
plateau of the resulting $am$ in $R_\mathrm{min}$. Furthermore, as the system never
undergoes symmetry breaking, we expect the constant $c_{1}$ to vanish
at all couplings in the infinite-volume limit.

\subsection{Numerical results}
\label{subsec:numerics}

\subsubsection{Beta-function}

The first task is the determination of the running of the scale. We measure
the correlator Eq.~(\ref{eq:2ptfct}) on lattices of size $156^{2}$ and $256^{2}$
for $\beta=0.1,0.2,\cdots,3.0$, generating $10^{5}$ configurations for each
setting, and use Eq.~(\ref{eq:correlat}) to extract $am(\beta)$.

\begin{figure}[ptb]\begin{center}
\includegraphics*[width=0.48\linewidth]{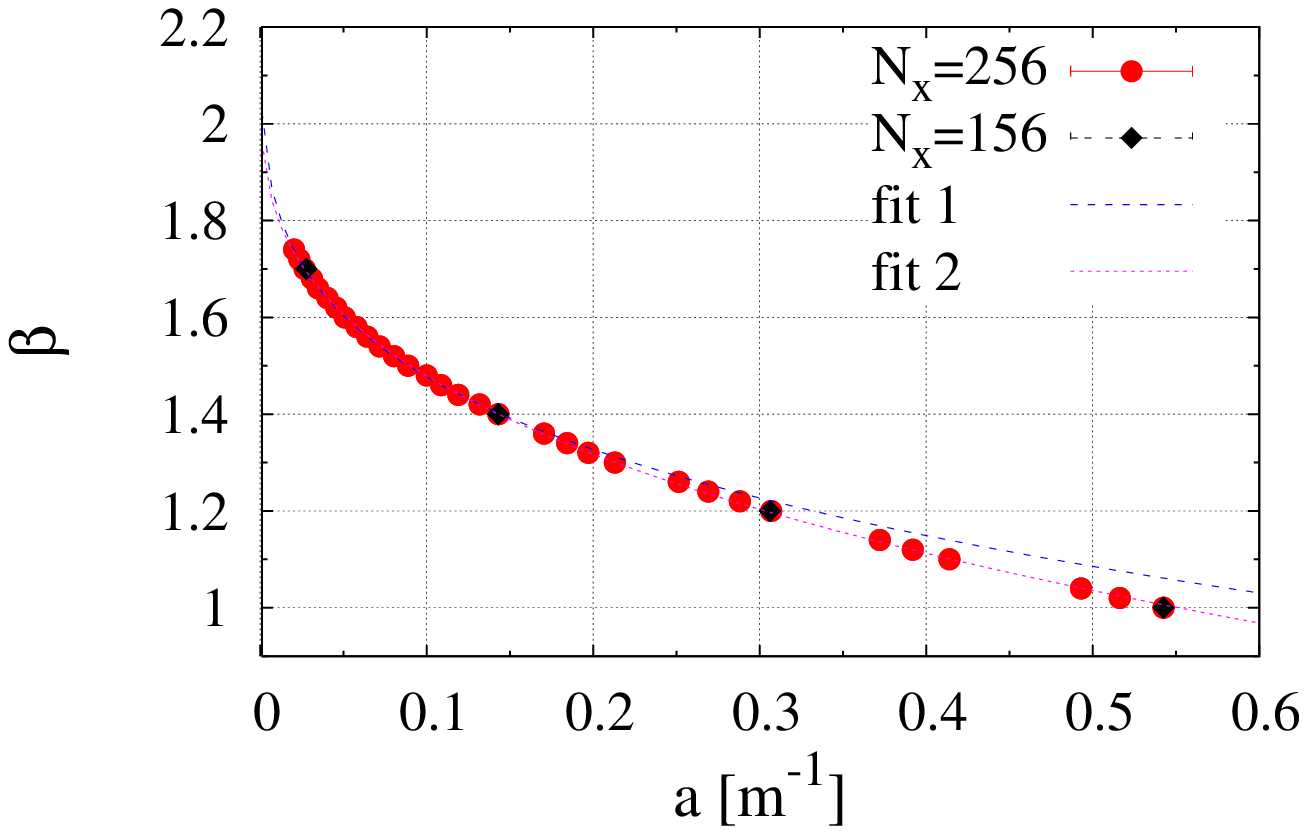}
\includegraphics*[width=0.48\linewidth]{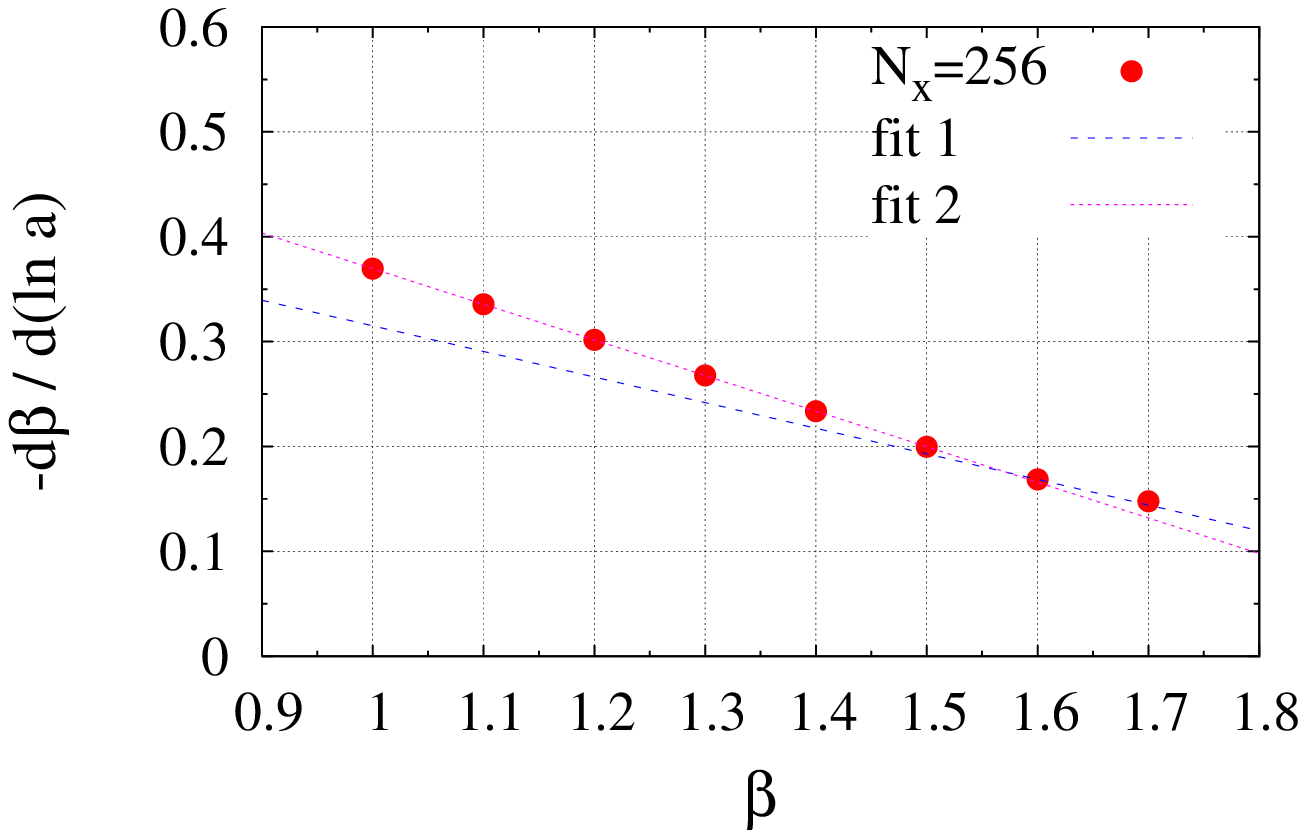}
\caption{
	\textit{Left:} the running coupling, computed on $156^{2}$ and $256^{2}$ lattices, with two
		parametrisations to the latter data, Eq.~(\ref{eq:running_fit}): fit 1 refers to
		$0.001 \leq am \leq0.08$, while fit 2 is done in $0.01 \leq am \leq0.6$. Error
		bars are much smaller than the symbol size; only some sample points are shown
		for the $156^{2}$ lattice.
	\textit{Right:} numerical results for the beta-function. Full circles indicate
		the final values used to set the physical scale.
}
\label{fig:run_coup}
\end{center}\end{figure}

With the system volumes at our disposal, we can fit the data to
Eq.~(\ref{eq:correlat}) and find a satisfactory plateau for the mass in the
region $\beta\lesssim1.8$: beyond that value, the correlation length $\xi=
1/(am)$ grows too much, making our system size insufficient
(Fig.~\ref{fig:run_coup}, left); correspondingly, beyond that value the constant $c_{1}$ in
Eq.~(\ref{eq:correlat}) ceases to be compatible with zero. On the other hand,
for couplings below $\beta\simeq1.0$, the correlator signal drops too fast,
i.e.~the lattice becomes too coarse to measure the lightest mass of the theory.

The evaluation of the coefficient Eq.~(\ref{eq:debeta_delogt}) is performed by
locally interpolating the measured data as
\eq
\beta(a) = g_{0}-g_{1}a^{\nu}\; \label{eq:running_fit}
\qe
and then differentiating the resulting curve; the interpolation is done on the
$256^{2}$ measurements, since a comparison with the $156^{2}$ data shows that
finite-size corrections are negligible. In this way we obtain the 
beta-function shown in Fig.~\ref{fig:run_coup} (right). The increasing difficulty in
fixing the scale as $\beta$ grows will be the main source of systematic error
in the following results.

\subsubsection{Thermodynamics}

The integrand in Eq.~(\ref{eq:presint}) is evaluated on lattices with spatial
sizes $N_{x}=156$ and $256$, and all time extents \mbox{$N_t=6,\cdots,18$} for
the finite-temperature part, plus the $N_{t}=N_{x}$ zero-temperature systems.
We measure the average energy per site at all couplings $\beta^{\prime}
\leq\beta$ in steps of $0.1$ and perform the integration evaluating the area
under the curve\footnote{The integral was done using a simple trapezoid
method. This gives satisfactory results since for the $\beta$ stepsize which
was used the curvature of $\ell(\beta)$ is very small.}, obtaining the
pressure for $\beta=1.0,1.1,\cdots,1.7$, corresponding to lattice spacings in
the range \mbox{$0.02 \lesssim am \lesssim 0.55$}.

For each combination of $N_{t}, N_{x}$ and $\beta$, $10^{5}$ measurements were
taken. The statistical error is estimated by binning the data in chunks of
$1000$, using Eq.~(\ref{eq:presint}) and averaging the resulting points for the
physical pressure: it turns out that the uncertainties are very small and
completely overcome by systematics.

\begin{figure}[ptb]\begin{center}
\includegraphics*[width=0.75\linewidth]{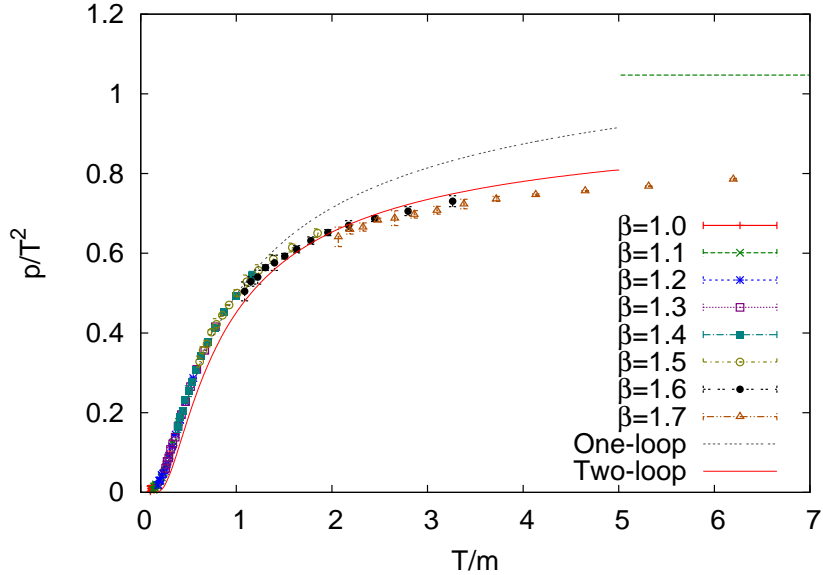}
\caption{The physical pressure $p(T/m)/T^{2}$ from $N_{x}=256$ systems at
	various lattice spacings. Each symbol represents a different $(\beta, N_{x})$.
	Error bars show the systematic error from both discretisation and finite-size
	effects. The lines represent analytical calculations at LO and NLO
	(Eqs.~(\ref{pren}) and (\ref{pren2loop}), respectively); the
	horizontal line marks the asymptotic value $\frac{\pi}{3}$.}
\label{fig:pressure}
\end{center}\end{figure}

The results for $p$ are plotted in Fig.~\ref{fig:pressure}, along with the
analytical expectations.
The slight differences between discretisations and system volumes are
used to assess the systematic errors; however, the neat agreement between the
various data sets is such that we consider both the thermodynamic and
continuum limits to be effectively reached.

\begin{figure}[ptb]\begin{center}
\includegraphics*[width=0.75\linewidth]{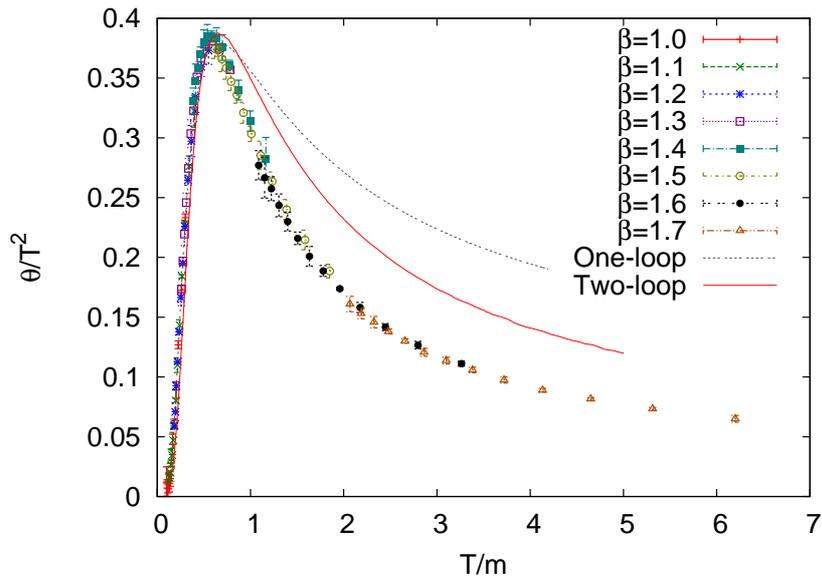}
\caption{Trace anomaly $\theta(T/m)/T^{2}$; symbols are as in Fig.~\ref{fig:pressure}.}
\label{fig:traceanomaly}
\end{center}\end{figure}

\begin{figure}[ptb]\begin{center}
\includegraphics*[width=0.75\linewidth]{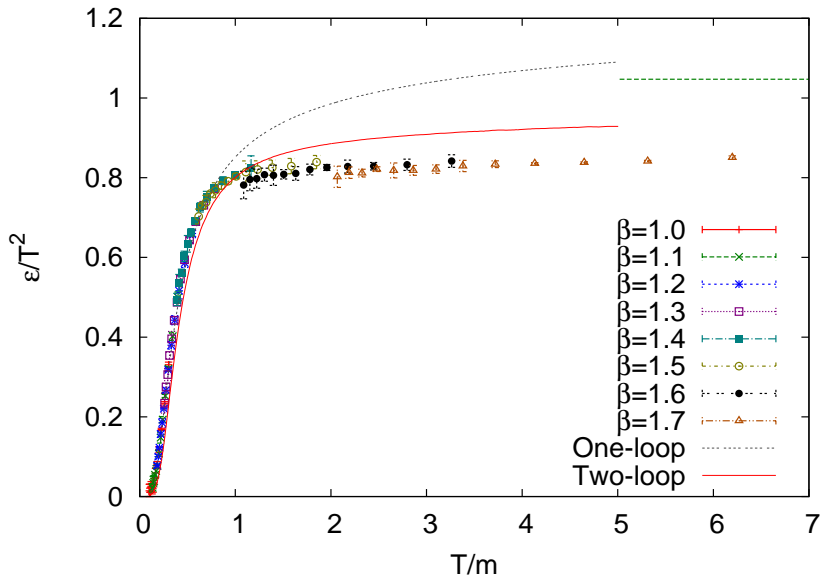}
\caption{Energy density $\epsilon(T/m)/T^{2}$; symbols are as in Fig.~\ref{fig:pressure}.}
\label{fig:energydensity}
\end{center}\end{figure}

Similar considerations hold for the trace anomaly, calculated using the same
input data through Eq.~(\ref{eq:tracanom}), and the energy density, obtained
with Eq.~(\ref{eq:en_density}). These are plotted in
Figs.~\ref{fig:traceanomaly} and \ref{fig:energydensity} respectively, along
with the analytical predictions.

\section{Comparison of analytical and lattice results}
\label{sec:comparison}

In this Section we compare the analytical results obtained in Sec.~\ref{sec:analytical} with the
lattice ones computed in Sec.~\ref{sec:montecarlo} for the case $N=3$. We organise this Section
by first discussing pressure, trace anomaly, and energy density;
then, we turn to two consequences of our study: the behaviour of the quasi-particle mass at
nonzero temperature and the role of instantons/calorons in the thermodynamics.

\bigskip

\emph{Pressure.} In Fig.~\ref{fig:pressure} the pressure is plotted as
function of $T$: we present the analytic curves at one- and two-loop order and
the lattice points for different values of $\beta.$ As one can deduce from the
plot, the one-loop approximation describes well the lattice data for small
temperatures. However, it overshoots them for large $T.$ The pressure at
two-loop level considerably improves the agreement of theory and lattice
simulations for high $T.$ Although a departure is still visible for increasing
temperature, we can conclude that there is a good and parameter-free agreement
between these two very different methods (the CJT formalism and lattice) of
evaluating the pressure of the system.

Notice that this finding could be expected by a closer inspection of the
properties of our analytic expressions: in fact, at one-loop level a good
agreement with lattice data is not achievable because of a wrong asymptotic
limit. Namely, in both cases the pressure approaches for high $T$ the limit of
a non-interacting gas of bosons, where each degree of freedom contributes as
$\pi T^{2}/6$. However, at one-loop level there are, at high $T$, $N$
active degrees of freedom, thus $p^{\text{(1-loop)}}(T\gg m)=N\pi T^{2}/6$ for
$T\rightarrow\infty$. This cannot be correct because the nonlinear constraint
in Eq.~(\ref{constraint}) eliminates one degree of freedom. On the contrary,
at the two-loop level one obtains the correct result $p^{\text{(2-loop)}}(T\gg
m)=(N-1)\pi T^{2}/6$ for $T\rightarrow\infty,$ which shows that only $N-1$
non-interacting bosons are present, as it should be. This fact explains the
much better agreement of the two-loop approximation at high temperature.

\bigskip

\emph{Trace anomaly.} In Fig.~\ref{fig:traceanomaly} we present a similar plot
of the function $\theta/T^{2}$. The apparent larger mismatch at high $T$
relies on the fact that $\theta=\epsilon-p=T^{3}\partial_{T}(p/T^{2})$: the
presence of a derivative enhances the disagreement for large $T.$ However, it is
very interesting to stress that both approaches show a form of the
function $\theta/T^{2}$ which looks very much like the one obtained in the
deconfined phase of Yang-Mills theory, e.g.~\cite{boyd,panero,borsanyi2012}. In fact, the
present two-dimensional model, which does not have any phase transition,
represents a simplified description of a gluonic gas above the critical
temperature. Note that also in lattice simulations of Yang-Mills theories the
maximum of the trace anomaly is reached above $T_{c}$ just as our
low-dimensional model does.

It is interesting to observe that $\theta$ grows linearly with $T$ for large enough $T$.
The lattice simulation shows that, for $T\gtrsim m$, the behavior $\theta \sim mT$ is realised:
this simple power law may point to interesting features of the present theory.
Also in four-dimensional Yang-Mills theory the behaviour of the trace anomaly was studied:
in Ref.~\cite{Pisarski:2006yk} a quadratic behavior has been found.
A similar finding has also been obtained in the analytical study of Ref.~\cite{giacosa_prd83},
whereas in Refs.~\cite{Giacosa:2007tga, Miller:2006hr} a linear rise of the trace anomaly was determined.
A deeper theoretical understanding of the behaviour of the trace anomaly would be
surely important: the linear growth seen here for the simple $O(3)$ model
may represent a suitable toy model for this purpose.

\bigskip

\emph{Energy density.} In Fig.~\ref{fig:energydensity} we present the plot
for the energy density. Analogous comments regarding the comparison of lattice
and analytic results hold. Moreover, here it is clearly visible that the
one-loop result overshoots the corresponding correct Boltzmann limit of $(N-1)$
active DOFs in the high-$T$ regime.

\bigskip

\begin{figure}[ptb]\begin{center}
\includegraphics*[width=0.69\linewidth]{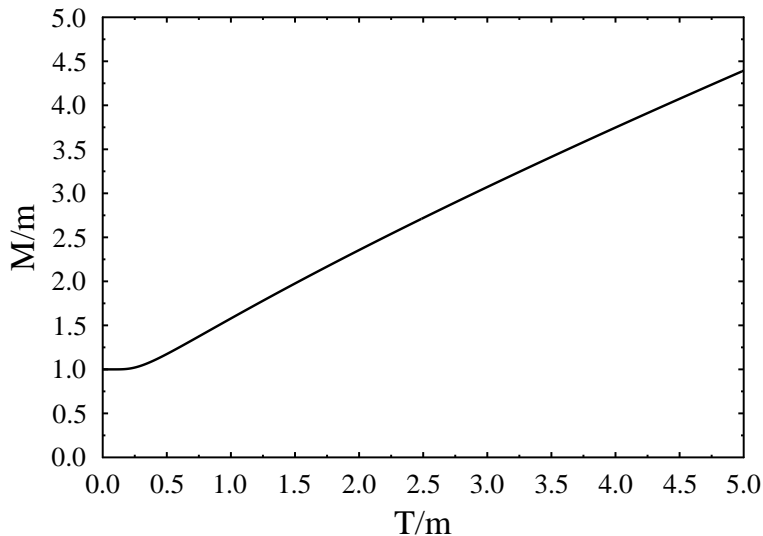}\caption{The quasi-particle
mass $M(T)$ as function of $T.$ For $T\rightarrow0$ the constant value
$M(0)=m$ is reached (dimensional transmutation). For high $T$ the perturbative
behaviour $M(T)\sim T/\ln(T)$ is realised.}
\label{fig:m}
\end{center}\end{figure}

\emph{Quasi-particle mass.} In the framework of our analytic approach we have
calculated the mass of the particles as function of $T.$ In Fig.~\ref{fig:m}
we plot the mass of the scalar particle $M(T)$ normalised to its vacuum
value $M(0)=m$ as function of the temperature (notice that the mass $M(T)$  is
the same for both the one-loop and the two-loop calculations). The function
$M(T)$ starts from a nonzero value at $T=0$ and then bends and reaches, as
expected by basic dimensional consideration, a straight line for large $T$.
This behaviour is reminiscent of that of the gluon mass in the deconfined phase
of Yang-Mills theories \cite{quasiparticle_1,quasiparticle_2,quasiparticle_3}. The model exhibits dimensional
transmutation just as QCD, meaning that at zero temperature there is a
nonvanishing mass gap, which is generated due to renormalisation of quantum
corrections. Note that the temperature at which the function $M(T)$ bends
corresponds roughly to the maximum of the trace anomaly. Conversely, at high
$T$ the temperature-dependence of the mass can be approximately parametrised
by $T/\log T$ in accordance with the aforementioned results concerning Yang-Mills theories.

\bigskip

\emph{Role of instantons.} The $O(3)$ model in $1+1$ dimensions exhibits
instantons in the vacuum and calorons at nonzero temperature. It is natural to
ask if and how these topological objects affect thermodynamics. To this
end, it should be stressed that the analytical method presented in Sec.~\ref{sec:analytical}
does not distinguish between trivial and nontrivial filed configurations,
neither in the vacuum nor at nonzero temperature. At a first sight, one may
then argue that no instanton effects are present in our model: in fact, the
method employed is valid for each $N$, while instantons are a peculiarity of
the $N=3$ case only. However, a closer inspection of our results is needed:
namely, the system --- as illustrated in Sec.~\ref{sec:analytical} --- shows
nonperturbative properties: the
vacuum is not realised by a constant value of the fields, such as
$\sigma=\phi\neq0$, rather by $\phi=0.$ Considering that the simple field
configurations $\sigma=0, \pi_i=0$ is obviously not allowed (it violates the
constraint), this means that rather complicated functions for $\sigma(x,\tau)$
and $\pi_{i}(x,\tau)$ with zero average are realised. For instance, an
ensemble of calorons and anticalorons fulfills the required properties: all
fields have zero average but nonzero contributions. We thus argue that the
role of topological objects is prominent, because it naturally explains the nontrivial
emergence of the vacuum $\left\langle \sigma\right\rangle =\left\langle
\pi_{i}\right\rangle =0.$

Indeed, in Ref.~\cite{warringalast} the $O(3)$ model has been compared to the
equivalent $CP^{1}$ model. In the case of $CP^{n-1}$ models a calculation for
the pressure was done only in the topologically trivial sector: in fact, in
that case it was possible to distinguish between the different topological
sectors of the theory. A comparison of the pressure of the $CP^{1}$ model with
that of the $O(3)$ model shows a large discrepancy, which was ascribed to
instanton/caloron effects by the authors of Ref.~\cite{warringalast}.
Obviously, this conclusion holds only under the assumption that the instanton
contributions are indeed included in the thermodynamical treatment of the $O(3)$ model.
As we argue above, this seems to be the case.

In this context, it is interesting to observe that the lattice data show
a linear growth of the trace anomaly with the temperature.
This feature implies a nonperturbative bag pressure which also grows linearly
with $T$:
in Ref.~\cite{Giacosa:2007tga} this linear behaviour of the trace anomaly
has been linked to a ground state dominated by caloron/anticaloron fluctuations.
Thus, a direct calculation of the instanton/caloron 
role in the $O(3)$ model would be very desirable from both
the analytic and the lattice sides. Moreover, the $O(3)$ calorons contain also
monopole constituents, just as Yang-Mills theories \cite{bruk2}. An estimate of the
contributions of these constituents to thermodynamics would also be very interesting.

\section{Summary}
\label{sec:conclusions}

In this work we have studied the thermodynamics of the $O(N=3)$ model in 1+1
dimensions using the CJT formalism and the auxiliary field method to one- and
two-loop orders. At the same time, we have performed a precise lattice
simulation for this system at nonzero temperature applying the integral method.

Our main results are displayed in Figs.~\ref{fig:pressure},
\ref{fig:traceanomaly} and \ref{fig:energydensity} where the theoretical
curves have been compared to the lattice calculation. It should be stressed that,
once the mass of the field in the vacuum is used as an energy unit, no free
parameters in the analytical calculation are present: the agreement shown in
the aforementioned figures confirms the power of the employed analytic
formalism. In particular, at one-loop level the results are very good for
small temperature, but not so at large $T$ because of an expected failure in
this energy domain. Namely, in this approximation a gas of free $N$ bosons is
realised in this limit, while only $N-1$ degrees of freedom exist due to the
nonlinear constraint entering the definition of the model. At two-loop
order this problem is solved and the correct asymptotic limit is reached. The
agreement of theory and lattice is therefore much better in the high-$T$
domain. Surely, theoretical improvement can still be done in order to achieve
an even better matching.

The $O(N)$ model is very interesting because it shares some properties with
Yang-Mills theory, such as trace anomaly and asymptotic freedom. In fact,
the form of the trace anomaly (see Fig.~\ref{fig:traceanomaly}) shows a
striking analogy to the more difficult four-dimensional Yang-Mills case. For
these reasons the lattice results presented here can be of interest for
testing other theoretical methods dealing with nonperturbative features of
quantum field theory. Namely, while going beyond the two-loop CJT results
presented in this work would be a very hard technical task, one could develop
new theoretical methods which involve new resummation schemes. Surely, a
better understanding of such methods in a model which shares many similarities
with Yang-Mills gauge systems can be of great use to improve our knowledge on
the latter.

A further interesting achievement has been the determination of the
temperature-dependent mass $M(T)$, see Fig.~\ref{fig:m}. Also in this case a
form which is reminiscent of that of the gluon mass
in the deconfined phase of Yang-Mills systems is
obtained: for small $T$ one has $M(T)\simeq m\neq0$ (dimensional
transmutation), while for large $T$ one has the expected perturbative growth
$M(T)\sim T/\ln T$.

For the particular choice $N=3$ used in this work the $O(3)$ model contains
also instantons. We have argued that these topological configurations,
although not explicitly included in our evaluation, are anyhow present because
they naturally generate the nontrivial vacuum of the system. Indeed, a further
study in this direction represents an interesting outlook both as an
analytic and as a lattice study of the $O(3)$ system.

One can also include in the present model a so-called $\theta$-term
which mimics the axial anomaly of QCD (see for instance \cite{Nogradi:2012dj}):
a thermodynamical investigation of this system
is feasible both analytically and numerically.

We also mention the possibility of a comparison between the results presented here and
an analysis based on a thermodynamic Bethe Ansatz adjusted to the model under study.

\bigskip

\section*{Acknowledgements}

The authors would like to thank Harmen Warringa for his precious help,
suggestions and discussions in the analytical part of this work. The authors 
thank also Dirk H.~Rischke, Owe Philipsen, Stefan Schramm, Rob Pisarski, Hendrik 
van Hees, Tomas Brauner and Ralf Hofmann for valuable discussions and a critical 
reading of the manuscript.
The numerical simulations presented here were
performed at the Center for Scientific Computing (CSC) at Frankfurt
University. This work has been partly supported through the Helmholtz
International Center for FAIR which is part of the Landes-Offensive zur
Entwicklung wissenschaftlich-\"{o}konomischer Exzellenz (LOEWE) of the state
of Hesse. F.G. acknowledges financial support form the Politechnical Society
of Frankfurt/M.~through an Educator fellowship.

\appendix
\section{Algorithms and autocorrelations}
\label{app:algorithms}
We now come to a study and comparison of the various
algorithms that can be implemented to investigate numerically the model,
Eq.~(\ref{eq:s_latt}). Aside from the ordinary, quite inefficient Metropolis
algorithm, which we use only for cross-checking purposes, we tested the
heatbath update method, the cluster method and the microcanonical overrelaxation.

In the following we measure and compare the most important indicator of an
algorithm's efficiency, namely the (integrated) autocorrelation time ${\tau
_{int}}$, which measures how many elementary update steps are needed to
generate a new, statistically uncorrelated configuration while exploring the
phase space of the theory through the Markov chain driven by the update
method. Furthermore, the final measure of the efficiency of an algorithm is
not ${\tauint}$ itself (measured in numbers of updates), but rather
${\tauintreal}$ (measured in seconds), the product of ${\tauint}$
and the average wallclock time it takes, on a given machine, to
perform a single update (which, in particular for cluster-based methods, may
largely depend on the coupling). All simulations presented in this work have
been done on individual cores of AMD Opteron 2427 (Istanbul) CPUs with a clock
rate of $2211$ MHz. No parallelisation has been employed.

In general, different autocorrelations are associated to different
observables. We focus on three quantities: two are of obvious physical
interest, corresponding to the energy- and ordering-like relevant scaling
operators, which are general for $O(N)$ models,
\eq
e = \frac{\Big\langle \sum_{\langle i,j \rangle} (1-\vec{s}_{i} \cdot\vec
{s}_{j}) \Big\rangle}{2N_{x}N_{t}} \;,\; m = \frac{\Big\langle \big| \sum_{\ell}
\vec{s}_{\ell} \big| \Big\rangle}{N_{x}N_{t}}\;\;. \label{eq:observables}
\qe
The third one is the volume-average of a single component of the $\vec{s}_{i}$
and gives a measure for ``isotropy-like'' correlations, i.e.~a memory of prior
global direction in internal space,
\eq
	i = \frac{\Big\langle \sum_{k} \vec{s}_{k} \cdot\hat{e}_{1}
		\Big\rangle}{N_{x}N_{t}} \;\;.
\qe
Its autocorrelation, which is technically interesting and can in fact become
quite large (where the system has large spatial correlations), is however for
the most part irrelevant from the quantum field theory perspective, which is
concerned primarily with $O(3)$-symmetric physics.

To estimate the autocorrelation, one can employ a binning analysis with bins
of variable size \cite{berg}, or use the jackknife method; here, we apply the
more refined technique proposed in \cite{Wolff:2003sm}. All autocorrelation
measurements have to be made after thermalisation is reached, which in the
present case is determined by inspection of the Monte Carlo time histories for
the observables defined above.

In the following we discuss the algorithms we implemented in terms of their
${\tauint}$ and ${\tauintreal}$, and present the mixed strategy
we adopted for all simulations reported in Subsection \ref{subsec:numerics}, which
ensures efficiency over the whole $\beta$-range of interest.

\vspace{0.5cm}
\noindent \textit{Heatbath} \\
The model, Eq.~(\ref{eq:s_latt}), is suited for the heatbath update. The new
vector on site $i$ is extracted randomly according to the equilibrium
probability distribution induced by the associated contribution to the action,
involving its four neighbours:
\eq
P(\vec{s}_{i}) \sim\exp\Big[ \beta\vec{s}_{i} \cdot\Bigg( \sum_{\hat{\nu}
=\pm\hat{e}_{j}} \vec{s}_{i+\hat{\nu}} \Bigg) \Big] \label{weight1}
\qe
In practice, it amounts to generating the angle between the sum vector and the
new $\vec{s}_{i}$ according to the above formula, and then uniformly choosing
a random azimuthal angle \cite{berg}.

\begin{figure}[ptb]\begin{center}
\includegraphics*[width=0.325\linewidth]{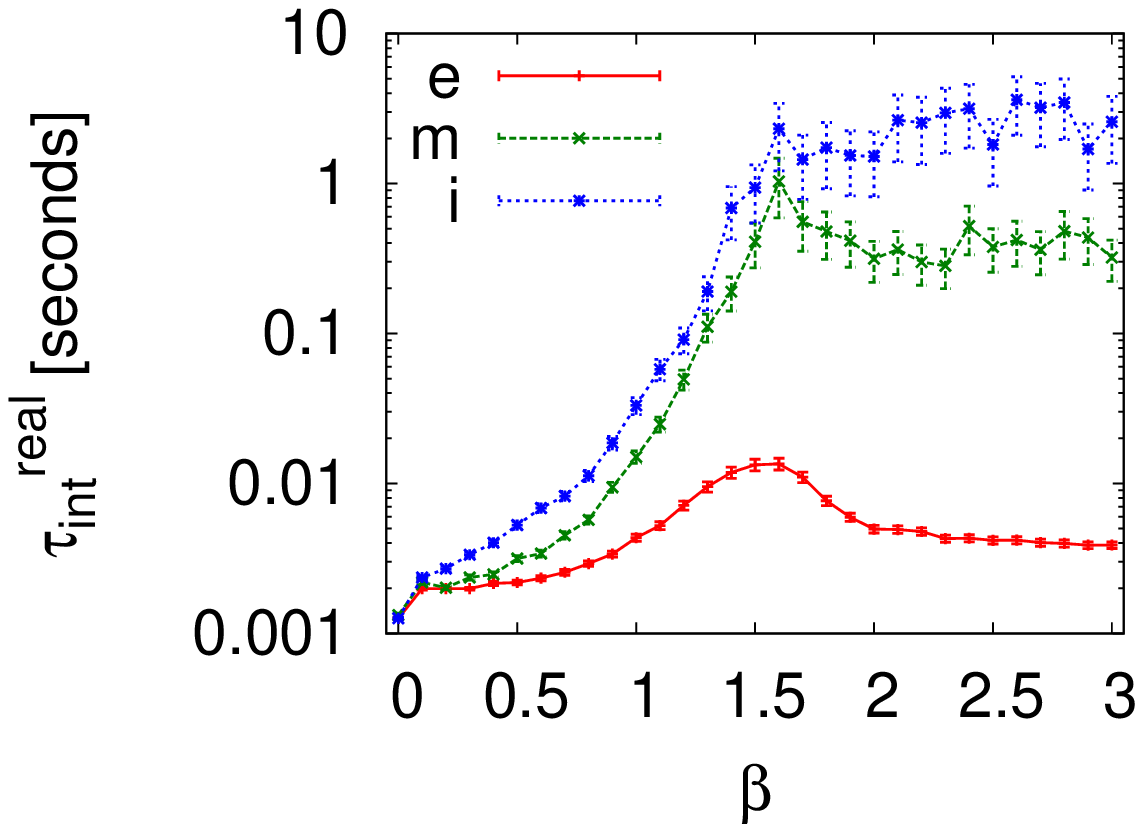}
\includegraphics*[width=0.325\linewidth]{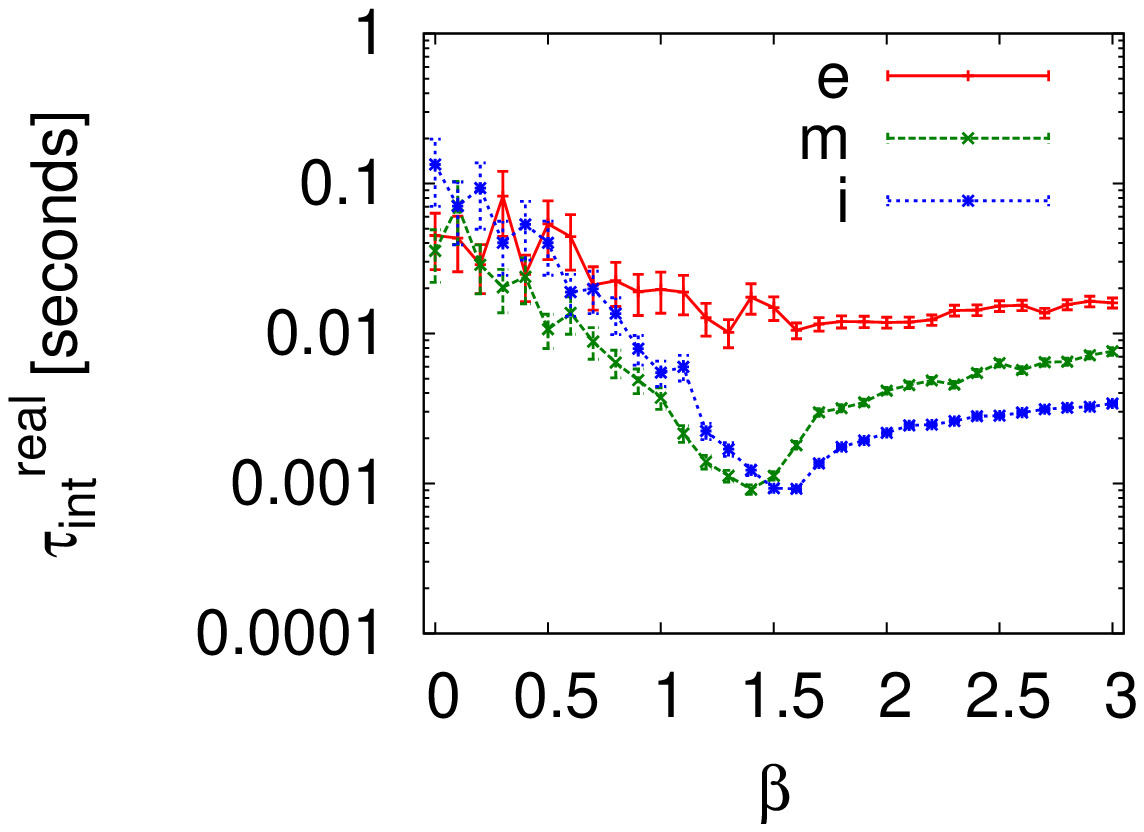}
\includegraphics*[width=0.325\linewidth]{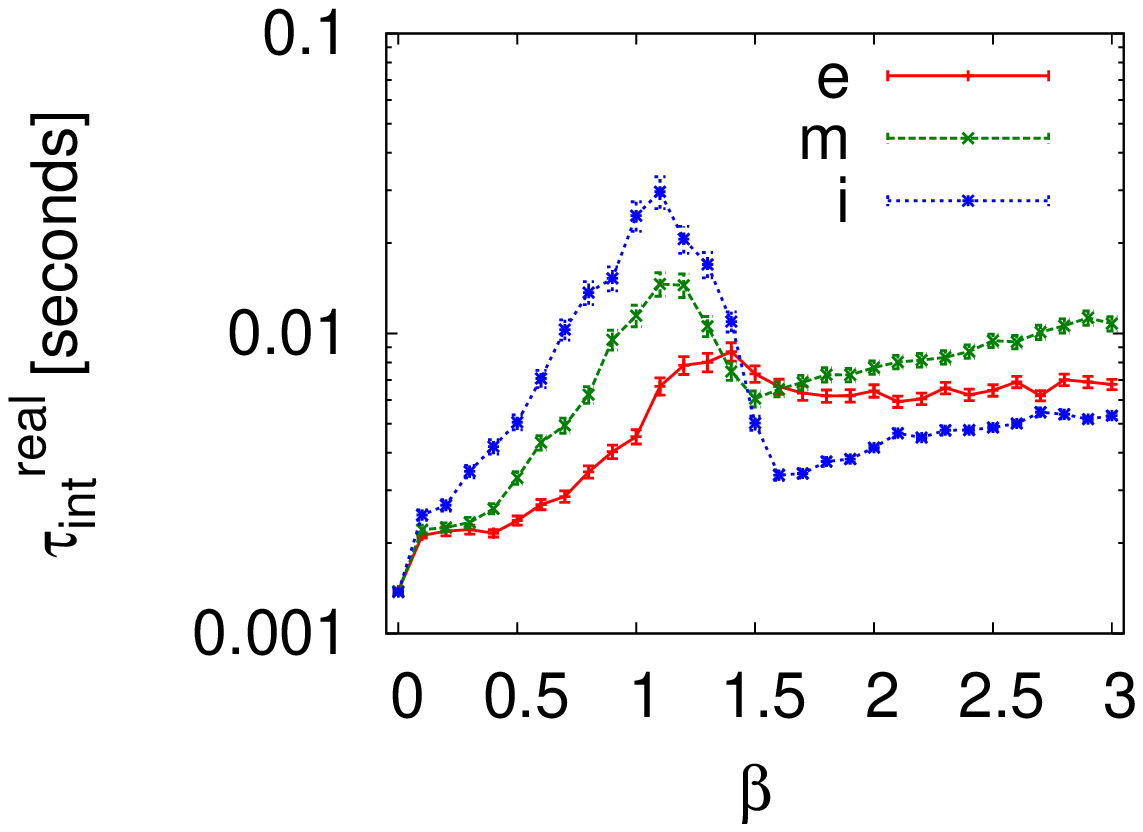}
\caption{Real-time ${\tauintreal}$ from simulations on a $64^{2}$
lattice, for heatbath (left), Wolff (middle) and mixed (right) update methods.
Note the semi-logarithmic scale.}
\label{fig:time}
\end{center}\end{figure}

All sites of the lattice are updated in a checkerboard fashion: this
constitutes an update step (``sweep''). In Fig.~\ref{fig:time} (left) the
behaviour of ${\tauintreal}$ for the three observables on a $64^{2}$
lattice is shown (we focus on this quantity instead of ${\tauint}$ to
make a meaningful comparison between different update methods, however for the
heatbath update the two are proportional in a $\beta$-independent way). It is
apparent that the method, while performing well for small couplings, suffers
from the long-range correlations that develop as $\beta\gtrsim1$, especially
for the ordering-like operators: there, alternative methods should be considered.

\vspace{0.5cm}
\noindent \textit{Cluster algorithm} \\
We then test the cluster update method, developed by Wolff \cite{Wolff:1989uk}
generalising the work of Swendsen and Wang \cite{Swendsen:1987ce} for the
Ising model. The idea is to build clusters of spins, reflecting to some extent
the structures, in the system, which carry the long-range correlations, and
update these clusters all at once. For our model, an update step consists of
the following operations: \renewcommand{\labelenumi}{\alph{enumi})}

\begin{enumerate}
\item Choose a random site $x_{i}$ and a random reflection plane $\pi$ with
surface normal $\hat{\vec{n}}$ (``seeding''). Reflect $\vec{s_{i}}$ across
$\pi$ and mark $x_{i}$.

\item Examine all nearest neighbours $\{x_{j}\}$ of $x_{i}$ and with
probability
\[
p=1-\exp\lbrace\min[0,2\beta(\hat{\vec{n}}\cdot\vec{s_{i}}) (\hat{\vec{n}
}\cdot\vec{s_{j}})]\rbrace
\]
reflect $\vec{s_{j}}$ across $\pi$ and mark $x_{j}$ as well.

\item Repeat step b) for all links connecting newly marked sites to unmarked
neighbours until the process stops.
\end{enumerate}

Clustering methods are generally understood to be free of critical slowing
down: thus, they are suitable for simulations in the vicinity of a critical
point. Our model does not exhibit a true transition in two dimensions,
nevertheless we implement the Wolff algorithm and investigate its efficiency,
since on any finite volume we see a strong ordering tendency at large enough couplings.

The average size of a cluster (therefore, the time needed perform a single
update) grows with $\beta$: indeed, we observe it increases by more than an
order of magnitude between $\beta=0.1$ and $\beta=3.0$. This is, however,
completely countered by a shortening of the autocorrelation as the clusters
span a larger portion of the system, so that, in terms of ${\tauintreal}$,
the efficiency of this algorithm in disordering the configuration
is generally better at larger couplings (Fig.~\ref{fig:time}, middle). In the
region of small $\beta$, the heatbath is still superior than this method, so
in the following we devise a mixed update that ensures small
autocorrelation over the whole range of couplings under study.

\vspace{0.5cm}
\noindent \textit{The mixed strategy} \\
Based on the findings presented so far, we expect that a judicious combination
of the two algorithms would perform well at all couplings. We then define from
now on an ``update sweep'' as the combination of a single heatbath sweep
throughout the lattice combined with ten cluster seedings.

Fig.~\ref{fig:time} (right) shows the resulting ${\tauintreal}$ on a
$64^{2}$ system: this update is indeed more efficient than either
heatbath or cluster alone over the whole $\beta$ range. Hence, we use
the mixed strategy for the thermodynamical investigation presented in
Subsection \ref{subsec:numerics}.

\begin{figure}[ptb]\begin{center}
\includegraphics*[width=0.46\linewidth]{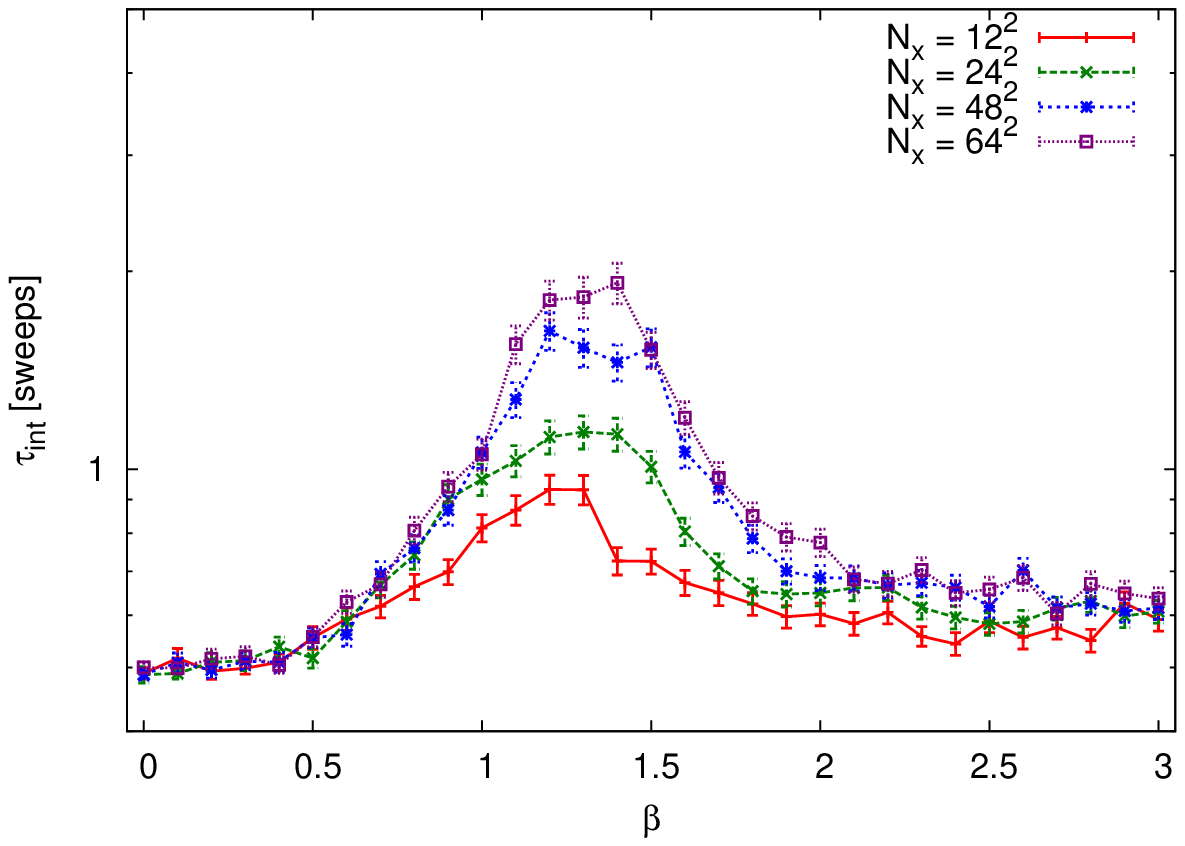}
\includegraphics*[width=0.46\linewidth]{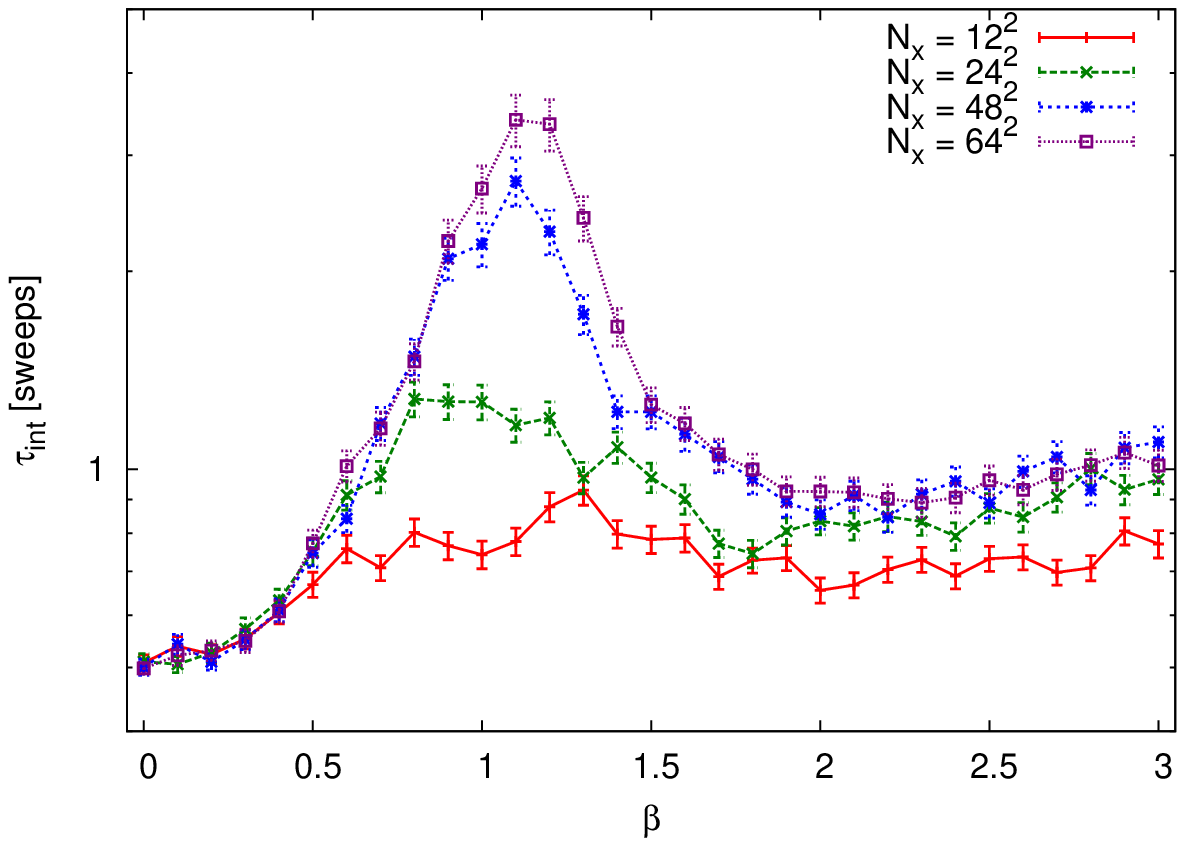}\caption{${\tau
_{int}}$ as a function of the coupling for the mixed update on square
systems of various sizes, for the $e$ (\textit{left}) and the $m$ (\textit{right}) observables.
Note the semi-logarithmic scale.
}
\label{fig:tint_mixed_volscaling}
\end{center}\end{figure}

We now turn to the question of the scaling of ${\tauint}$ with the system
volume for the mixed update: this is useful for a proper tuning of the Monte
Carlo data taking based on the size of the lattice under study. We measure
${\tauint}$ in units of sweeps, as defined above, on various square
systems with size up to $N_{x}=64$ for energy- and ordering-like observables;
these are plotted in Fig.~\ref{fig:tint_mixed_volscaling}. The latter seems
always to be a bit slower to decorrelate, so we base our tuning on its
behaviour. It turns out that the height peak of ${\tauint}(\beta)$,
located roughly at $\beta\sim1.2$, can be described as a function of the
system size with a power law:
\eq
{\tau_\mathrm{int}^*}
(N_{x}) \simeq a N_{x}^{2b} \;\;,\;
a=0.13(5)\;,\;b=0.38(6)\;,
\qe
this implies the upper bounds ${\tau_\mathrm{int}^*}(156)\sim6$ and 
${\tau_\mathrm{int}^*}(256)\sim9$; to be on the safe side, in collecting the data of
Subsection \ref{subsec:numerics}, we perform respectively 14 and 21 sweeps
between measurements.

\vspace{0.5cm}
\noindent \textit{Metropolis and overrelaxation} \\
For the purpose of cross-checking, we also implemented a standard Metropolis
update, where a proposed new value of $\vec{s}_{i}$ is accepted with
probability \mbox{$p=\textrm{min}[1,\exp(-\Delta S)]$}, with $\Delta S$ the
corresponding change in the action. The associated autocorrelation is
qualitatively similar to the heatbath procedure, but it is about an order of
magniture larger than the latter.

We also investigated whether the algorithm's efficiency can be improved with
microcanonical overrelaxation, i.e.~a random rotation of $\vec{s}_{i}$ about
the sum of the neighbours' spins so to achieve maximum disordering of the
configuration while exactly preserving the energy and the Metropolis
acceptance. While this brings indeed the Metropolis update to an efficiency
comparable to the heatbath in the region $\beta\gtrsim1$, we do not see any
gain at lower couplings: therefore, we refrain from exploring this avenue any further.

\end{document}